\documentclass[preprint,12pt]{elsarticle}

\usepackage{graphicx}
\usepackage{mathtools}
\usepackage{amsmath}
\usepackage{amssymb}
\usepackage{bm}
\usepackage{parskip}
\usepackage{hyperref}
\usepackage{xcolor}
\usepackage{soul}
\usepackage{longtable}
\biboptions{numbers}

\newcommand{\ifib}[1]{I_{#1,\mathrm{eff}}}
\newcommand{\trace}[1]{\mathrm{\,tr}\bm{#1}}

\usepackage{geometry}
\hypersetup{pdfauthor={Nishan Parvez, Prashant K. Purohit, Wensi Wu}}

\begin{document}

\begin{frontmatter}
  \title{An Octahedral Fibrous Constitutive Model for Heart Valve Mechanics and Function}
  \author[chop1]{Nishan Parvez}
  \author[upenn]{Prashant K. Purohit\corref{coor}}
  \author[upenn,chop1,chop2]{Wensi Wu\corref{coor}}

  \cortext[coor]{Corresponding authors. Email addresses:
    purohit@engineering.upenn.edu (Prashant K. Purohit);
  wensiwu@engineering.upenn.edu (Wensi Wu).}

  \affiliation[chop1]{
    organization={Department of Pediatric Cardiology, Children's Hospital of Philadelphia},
    city={Philadelphia},
    postcode={19104},
    state={Pennsylvania},
    country={USA}
  }

  \affiliation[upenn]{
    organization={Department of Mechanical Engineering and Applied Mechanics, University of Pennsylvania},
    city={Philadelphia},
    postcode={19104},
    state={Pennsylvania},
    country={USA}
  }

  \affiliation[chop2]{
    organization={Cardiovascular Institute, Children's Hospital of Philadelphia},
    city={Philadelphia},
    postcode={19104},
    state={Pennsylvania},
    country={USA}
  }

  \begin{abstract}
    Fibrous soft tissues derive their nonlinear mechanical response from networks of extracellular matrix fibers. The organization of fibers and their load-bearing characteristics give rise to strain stiffening, the reverse Poynting effect, and anisotropic mechanical behavior. Motivated by these coupled features, we develop an anisotropic hyperelastic model for fibrous biological tissues that accounts for the contribution of the fiber network under both tensile and compressive deformation in an asymmetric fashion. The model uses an octahedral arrangement of fibers embedded in a ground matrix, allowing anisotropy, fiber volume fraction, and pre-stretch to influence the material response. We calibrate the model to experimental data for mitral valve leaflets using an inverse finite element approach that is coupled to automatic differentiation to facilitate efficient parameter calibration. The calibrated model reproduces responses across biaxial, pure-shear, and constrained-uniaxial tests, while separate numerical tests demonstrate its ability to represent substantial volume reduction and the reverse Poynting effect. Using the calibrated model, we investigate how anisotropy and fiber reorientation affect valve deformation under physiological loading. In the idealized valve, the reference fiber orientation, which provided greater radial compliance, yielded complete leaflet coaptation. Prescribed local fiber reorientation redistributed leaflet stress but did not compromise coaptation. Swapping the preferred material directions throughout the leaflets, by contrast, prevented closure in the simulated valve. Finally, we examine how fiber pre-stretch and collagen volume fraction affect leaflet strain and contact, finding a stronger response to changes in pre-stretch over the parameter ranges studied.
  \end{abstract}
\end{frontmatter}

%\tableofcontents
\clearpage
\section*{List of symbols}
\addcontentsline{toc}{section}{List of symbols}
\noindent Subscripts \(a\), \(d\), and \(f\) refer to axial, diagonal, and an arbitrary individual fiber. Indices \(i=1,2,3\), e.g., in \(\Box_{a_i}\), and \(j=1,2,3,4\), e.g., in \(\Box_{d_j}\), enumerate the axial and diagonal fiber directions.

\begin{longtable}{@{}p{0.28\linewidth}p{0.67\linewidth}@{}}
  \textbf{Symbol} & \textbf{Meaning} \\
  \hline
  \endfirsthead
  \textbf{Symbol} & \textbf{Meaning} \\
  \hline
  \endhead
  \(\bm F,\bm C,J\) & Deformation gradient, right Cauchy--Green deformation tensor \(\bm C=\bm F^{\mathrm T}\bm F\), and volume ratio \(J=\det\bm F\). \\
  \(\tilde{v}_f,\tilde{a}_i,\tilde{d}_j\) & Unit vectors along generic, axial, and diagonal fiber directions in reference configuration. \\
  \(L_f,L_{a_i},L_{d_j}\) & Generic, axial, and diagonal reference fiber lengths; \(L_{d_j}=L_d\) for all \(j\).\\
  \(A_f,V_o\) & Fiber cross-sectional area and initial unit-cell volume. \\
  \(\alpha_i\) & Direction cosine, \(L_{a_i}/L_d\). \\
  \(\phi_f\) & Fiber volume fraction; the matrix volume fraction is \(1-\phi_f\). \\
  \(E_f,E_a,E_d\) & Generic, axial, and diagonal fiber stiffness parameters. \\
  \(m_f,m_a,m_d\) & Generic, axial, and diagonal fiber stiffening exponents. \\
  \(E_n,\nu\) & Young's modulus and Poisson's ratio of the Neo-Hookean ground matrix. \\
  \(\lambda_f\), \(\lambda_o\) & Fiber stretch, pre-stretch.\\
  \(I_f, \ifib{f}\) & Squared fiber stretch, \(I_f=\lambda_f^2\). This is the fourth invariant of \(\bm{C}\) associated with the fiber direction. The effective squared fiber stretch with pre-stretch is denoted as \(\ifib{f} = \lambda_o^2 \lambda_f^2 \). \\
  \hline
  \(\Psi_f, \Psi_{\mathrm{fs}}, \Psi_g, \Psi\) & Strain energy of an individual fiber, fiber system, ground matrix, and their sum. \\
  \(\bm{S},\bm{S}_{\mathrm{fs}},\bm{S}_g\) & Total, fiber-system, and matrix second Piola--Kirchhoff stresses.\\
  \(\mathcal C_{\mathrm{fs}}\) & Fourth-order material tangent from the fiber system. \\
  \(\sigma\) & Cauchy stress: \(\sigma_{xx}\) and \(\sigma_{yy}\) are normal components in simple shear, \(\sigma_{xy}\) is the shear stress, and \(\sigma_1\) is the first principal stress.\\
  \hline
  \(\Psi_{\mathrm{c}}\) & Strain-energy density of the chordae tendineae material (in this work: compressible Ogden model).\\
  \(\lambda_1,\lambda_2,\lambda_3\) & Principal stretches used to define the Ogden model. \\
  \(c_p,c_i,m_i,N\) & Volumetric penalty coefficient, Ogden coefficients, exponents, and number of Ogden terms. \\
  \hline
  \(k_c,p,r_{\mathrm{in}},r_{\mathrm{out}}\) & Contact strength, exponent, and inner and outer contact ranges. \\
\end{longtable}

\newpage
\section{Introduction}
Soft fibrous materials functioning as load-bearing components of biological tissues and scaffolds of organic matter are ubiquitous in nature~\cite{chawlaFibrousMaterials2016, picuNetworkMaterialsStructure2022, boalMechanicsCell2012}. The mechanical behavior of these materials arises from the collective deformation of fibers connected in disordered or partially aligned networks. Under load, the fibers in such networks can rotate, stretch, bend, and reorganize in a non-affine manner, giving rise to nonlinear stiffening and non-classical shear-induced normal stress effects such as the reverse Poynting effect~\cite{janmeyNegativeNormalStress2007, licupStressControlsMechanics2015, islamMorphologyMechanicsFungal2017, destradeDominantNegativePoynting2015}. The fiber network architecture also determines the degree of anisotropy, which depends on tissue type, physiological function, and the conditions under which the fiber network forms. Heart valve leaflets provide a representative example: their collagen fibers are preferentially arranged along the so-called circumferential and radial directions, leading to markedly different material responses along these axes~\cite{rauschMechanicsMitralValve2013a, meadorDetailedMechanicalMicrostructural2020, sadeghiniaBiomechanicsMitralValve2022, rossBiaxialMechanicalCharacterizations2019}. Similar structure-function relations arise in blood vessels, tendons, ligaments, and other fibrous tissues, where fiber orientation and dispersion govern the macroscopic response~\cite{gasserHyperelasticModellingArterial2006, sharabiStructuralMotifsSoft2025, sharabiStructuralMechanismsSoft2022, picuNetworkMaterialsStructure2022}. These structural features are critical to tissue function and motivate constitutive models that can capture anisotropic fibrous mechanics under physiological loading.

Theoretical modeling of fibrous materials has historically followed two broad approaches: discrete material models and continuum constitutive models~\cite{stracuzziRiskyInterpretationsLength2022}. Discrete fiber network models explicitly represent fibers and their interactions through crosslinks, contacts, or filament-level mechanics. Finite element analysis (FEA) and related methods have been used to study such networks of fibers, e.g., collagen, revealing how properties such as fiber alignment, connectivity, and fiber bending control non-affine deformation and strain stiffening~\cite{islamEffectNetworkArchitecture2018, broederszCriticalityIsostaticityFiber2012, hutchinsonStructuralPerformancePeriodic2006, parvezStiffeningMechanismsStochastic2023, picuPoissonsContractionFiber2018, amjadStrengthStiffnessNetwork2025}. However, the computational cost of these models limits their direct application to small systems and idealized loading conditions. As such, continuum constitutive models provide a more practical route for FEA of complex geometries and boundary value problems~\cite{broederszCriticalityIsostaticityFiber2012, hermannHomogenizationRateindependentElastoplastic2025, pouliosHomogenizationLongFiber2016, rainaHomogenizationApproachNonwoven2014, chagnonHyperelasticEnergyDensities2015, picuConstitutiveModelsRandom2021, stracuzziRiskyInterpretationsLength2022}. Although these models represent fiber-level kinematics through effective constitutive assumptions rather than by explicitly resolving deformation of the microscopic fiber elements, they remain standard in FEA of fibrous tissue behavior. Consequently, their predictive capability depends strongly on how fiber-network contributions are incorporated into the constitutive law.

A central assumption in many existing constitutive models concerns the mechanical role of fibers under compression. The well-established Holzapfel-Gasser-Ogden (HGO) model, for example, captures anisotropic behavior through a strain-energy function that combines an isotropic ground matrix with distributed fiber contributions~\cite{holzapfelNewConstitutiveFramework2000, gasserHyperelasticModellingArterial2006, latorreTensioncompressionSwitchGasser2016, nikpasandHybridMicrostructuralContinuumMultiscale2021}. The fibers are assumed to have a Fung-type exponential strain energy function described generically as \(W \sim \left(e^{Q} - 1\right)\), where \(Q\) is a function of the fiber strain. Similar ideas appear in transversely isotropic formulations developed for heart valve tissues, e.g.,~\cite{wuAnisotropicConstitutiveModel2018}. Two other common forms are polynomial and power law based development in-line with Rivlin and Ogden hyperelastic models but operating on fiber stretch~\cite{chagnonHyperelasticEnergyDensities2015}. In most cases, however, fibers are assumed to contribute to the overall strain energy only when in tension. This is motivated by the buckling of, generally tortuous, fibrils in biological tissue under compression~\cite{peacockBucklingTorsionalInstabilities2020, domaschkeRandomAuxeticsBuckling2019a}. Although this treatment is effective in many applications, it leaves important network mechanisms unresolved. Fibers may not immediately buckle under compression; they may also bend, rotate, and reorganize because of their finite bending stiffness, surrounding matrix elements, and orientation relative to the loading direction. Discrete network studies show that deformation can remain bending dominated up to moderately large strains~\cite{parvezStiffeningMechanismsStochastic2023, picuPoissonsContractionFiber2018, licupElasticRegimesSubisostatic2016}. Dry fibrous materials, though far more compliant than in tension, also sustain significant compressive load before compaction~\cite{islamEffectNetworkArchitecture2018}. Similarly, fluid efflux during tensile loading can lead to volume reduction in soft tissues~\cite{ehretInversePoroelasticityFundamental2017}. Therefore, the incompressibility assumption in various hyperelastic models may also be inadequate or fail to capture actual material behavior. These observations point to the need for a general constitutive model that accounts for fiber-network contributions under both tensile and compressive deformation while representing anisotropic fiber orientation and nonlinear stiffening in a compressible material framework. To address this need, we developed an anisotropic hyperelastic constitutive model, termed the octahedral fibrous model, that incorporates these aspects in an anisotropic setting for soft fibrous materials.

In defining a strain-energy density function from a continuum mechanics viewpoint, we propose to model the effective micromechanical features and properties of a complex material system using a cuboid unit cell (equivalently - an octahedral configuration). We model a representative volume element of the material as a network of 14 `fibers' organized along the centerlines (hereafter referred to as axial directions) and diagonals of a cuboid or along the edges and inner diagonals of an octahedron. This particular fiber placement is motivated by a model proposed in~\cite{parvezStiffeningMechanismsStochastic2023} where it was postulated that exponential stiffening in fibrous materials is a consequence of geometric nonlinearities associated with deformation of fibers in an equivalent unit cell. In contrast to a discrete collagen fiber with finite bending rigidity, `fibers' in our model should be interpreted as equivalent constructs modeled as nonlinear springs that capture the effective material response. The strain energy function describing their deformation is a power-law function with a logarithmic compression penalty that has a smooth transition from tensile to compressive regime, first proposed in~\cite{markertGeneralPolyconvexStrainenergy2005}. Conceptually, the axial directions represent the tissue's preferred fiber orientations. The diagonal directions provide additional paths for load transfer and coupling between the primary axes. Geometrically, they resemble the oblique segments of force chain fibers that emerge in discrete fiber networks~\cite{mainakForceChain22, elejueneForceChain2025, parvezStiffeningMechanismsStochastic2023}. The overall stress response of the material model in any given direction is the algebraic sum of the contributions of the various fibers. The relative anisotropy is therefore controlled via the material properties of the fibers and the relative orientation of the diagonal fibers. Other works have also used a similar unit cell with differences in the energy function and nature of the fiber/chain, for example, in~\cite{tutwilerRuptureBloodClots2020, bischoffMicrostructurallyBasedOrthotropic2002, kuhlContinuumModelRemodeling2007, buganzatepoleStretchingSkinPhysiological2012, jiaMicromechanicalModelGrowth2019}. The proposed model combines tensile and compressive fiber response, fiber pre-stretch, and material anisotropy in a single framework.

To demonstrate the model's applicability to highly nonlinear anisotropic biological soft tissues, we use it to analyze the deformation and closure of the mitral valve. The material parameters are calibrated to experimental data from porcine mitral valve tissue using an automatic differentiation (AD)-assisted finite element model updating (FEMU) technique. We first simulate leaflet coaptation in an idealized mitral valve geometry with the calibrated model. We then examine the effects of fiber reorientation on mitral valve closure. In the model, swapping the preferred material directions throughout the leaflets impairs coaptation. We further observe stress concentration in response to localized fiber reorientation. Additionally, we investigate the effects of fiber pre-stretch and volume fraction. We find that decreasing pre-stretch leads to billowing, which can be partially compensated by an increased volume fraction. These simulations illustrate how changes in leaflet material properties can affect coaptation in the valve model.

Overall, in this work, we introduce an anisotropic material model that is applicable to soft fibrous materials. The presentation is organized as follows: first, we present the material model and introduce the relevant parameters. We then calibrate the material parameters to five sets of experimental data simultaneously. Next, we comment on several representative properties that can be modeled efficiently with the proposed material model, namely, strain stiffening, volume reduction, and the reverse Poynting effect. Finally, we present a concise case study focusing on the deformation of the mitral heart valve and explore the role of various parameters such as fiber orientation, fiber pre-stretch, and collagen volume fraction.

\section{Material Model and Methods}
\subsection{Strain Energy Density Function}

\begin{figure}
  \centering
  \includegraphics[width=0.8\linewidth]{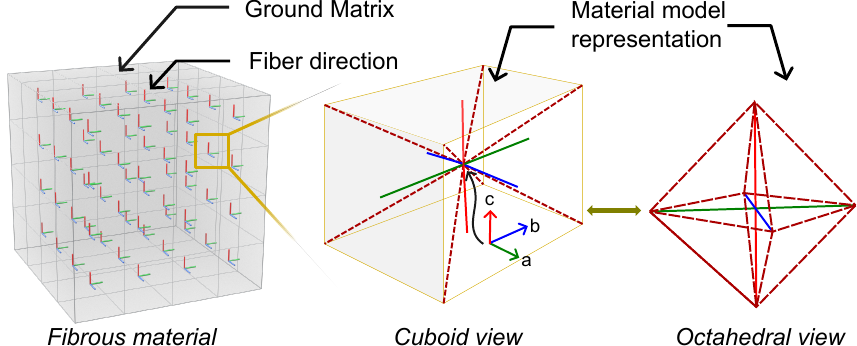}
  \caption{Unit cell representation of the proposed material model. The fibers are organized in an octahedral fashion within a continuum matrix. The primary axes of the unit cell are aligned with the fiber directions in an orthotropic fibrous material.}
  \label{fig:unit_cell_representation}
\end{figure}

The proposed material model contains two types of fibers: three center-to-face (axial) fibers, and four center-to-corner (diagonal) fibers. The fibers in the opposite directions complete the full set of this 14-fiber architecture. The fiber arrangements are shown in Figure~\ref{fig:unit_cell_representation}. The axial fibers are aligned with the primary material axes, e.g., the fibers in an orthotropic fibrous material, while the diagonal fibers span the oblique directions within the unit cell. Due to symmetry, the formulation below uses only three axial and four diagonal directions, with a prefactor of 2. We characterize the fibers using an energy function expressed in terms of the fourth invariant of the right Cauchy--Green deformation tensor, \(I_f = \bm{C}:(\tilde{v}_f \otimes \tilde{v}_f) = \lambda_f^2\). Here, \(\bm{C}\) is the right Cauchy--Green deformation tensor, \(\tilde{v}_f\) is a unit vector describing the direction of a fiber, and \(\lambda_f\) is the axial stretch of the fiber. The specific fiber directions and their associated invariants are defined in~\ref{sec:SI_fiber_dirs}.

Fibers in biological tissue such as heart valves are often pre-stretched in vivo~\cite{rauschMechanicsMitralValve2013a, aggarwalInvivoHeterogeneousFunctional2016, leeInVivoFunctionMitral2017,oomenGrowthRemodelingOpposing2018}. We incorporate a multiplicative pre-stretch and define an effective invariant to model such pre-stretches in the fibers. Let \(I_{a_i}\) and \(I_{d_j}\) denote the squared stretches of axial fiber \(i\) and diagonal fiber \(j\), respectively. The effective invariants are written as:
\begin{equation}
  \ifib{a_i} = \lambda_o^{2} I_{a_i}
  \qquad
  \ifib{d_j} = \lambda_o^{2} I_{d_j}
  \label{eq:prestretch_invariants}
\end{equation}

where \(\lambda_o>0\) represents the pre-stretch in the fibers. With \(A_f\) and \(L_f\) denoting the cross-sectional area and length of the fiber, respectively, we utilize the following strain energy for an individual fiber:

\begin{equation}\label{eq:fib_energy}
  \Psi_f = \frac{E_f A_f L_f}{2m_f-1} \left( \frac{\ifib{f}^{m_f}}{2m_f} - \frac{1}{2}\log\ifib{f} \right) + \Psi_o
\end{equation}

In the energy function, \(m_f\) is a material parameter that controls the nonlinear stiffening behavior of the fiber in tension. \(\Psi_o\) is an additive reference-energy constant. We set it to zero for the remainder of this work, since it does not affect the stress or tangent. This particular form of fiber energy function was first proposed by Markert et al.~\cite{markertGeneralPolyconvexStrainenergy2005}. Related invariant-based strain-energy functions for collagen networks were later proposed by Anssari-Benam et al.~\cite{anssariBenamSpecializedStrainEnergy2020}. The power-law term in Eq.~\ref{eq:fib_energy} dominates in tension, whereas the logarithmic term dominates in compression, leading to an asymmetric response of the fiber under tension and compression. The asymmetric fiber contribution distinguishes the proposed formulation from constitutive models that exhibit tension--compression symmetry in their fiber-related strain energy~\cite{wuAnisotropicConstitutiveModel2018,guo2007mechanical,guccione1991passive,may1998constitutive,humphrey1987new}. Overall, there is a smooth transition from tension to compression for the individual-fiber energy with a minimum at \(\ifib{f} = 1\). The fiber-system strain-energy density is obtained by summing the individual-fiber energies and dividing by the reference unit-cell volume:

\begin{equation}
  \begin{aligned}
    \Psi_{\mathrm{fs}} &= \frac{2}{V_o}\Bigg[\sum_{i=1}^{3} \frac{E_a L_{a_i} A_f}{2m_a-1}
      \left(\frac{\ifib{a_i}^{m_a}}{2m_a} - \frac{1}{2}\log\ifib{a_i}\right)\\
      &\qquad + \sum_{j=1}^{4} \frac{E_d A_f L_{d_j}}{2m_d-1}
      \left(\frac{\ifib{d_j}^{m_d}}{2m_d} - \frac{1}{2}\log\ifib{d_j}\right)
    \Bigg]
  \end{aligned}
  \label{eq:fiber_system_energy}
\end{equation}

Here, the subscripts \(a\) and \(d\) indicate axial and diagonal fiber material parameters and \(V_o\) is the initial volume of the unit cell. The prefactor of 2 is due to the fact that each fiber contribution appears twice in the unit cell. If different pre-stretches along the axial directions are desired, the sum of axial fiber contributions in the energy function may be expanded with different \(\lambda_o\) (through \(\ifib{a_i}\)) along these directions with appropriate modifications to the diagonal fibers.

We set all fibers to have the same cross-sectional area, \(A_f\). The lengths of the axial fibers are allowed to change independently. The four diagonal fibers have the same reference length, \(L_{d_j}=L_d\) for \(j=1,2,3,4\), with \(L_d^2 = L_{a_1}^2 + L_{a_2}^2 + L_{a_3}^2\). This relationship ensures geometric compatibility within the unit cell and leads to direction cosines, \(\alpha_i = L_{a_i} / L_d\). The volume fraction of fibers in a unit cell, \(\phi_f\), is defined as:
\begin{equation} % convert it to Ld and alpha
  \phi_f = \frac{2 A_f}{V_o}(L_{a_1} + L_{a_2} + L_{a_3} + 4 L_d) = \frac{2 A_f L_d}{V_o}(\alpha_1 + \alpha_2 + \alpha_3 + 4)
\end{equation}

We replace \(V_o\) with \(2 A_f L_d (\alpha_1 + \alpha_2 + \alpha_3 + 4)/\phi_f\) and \(L_{a_i} = \alpha_i L_d\) to rewrite \(\Psi_{\mathrm{fs}}\) as:
\begin{equation}
  \begin{aligned}
    \Psi_{\mathrm{fs}} &= \frac{\phi_f}{(\alpha_1 + \alpha_2 + \alpha_3 + 4)}\Bigg[\sum_{i=1}^{3} \frac{E_a \alpha_i}{2m_a-1}
      \left(\frac{\ifib{a_i}^{m_a}}{2m_a} - \frac{1}{2}\log\ifib{a_i} \right)
      +\\
      &\qquad\sum_{j=1}^{4}
      \frac{E_d}{2m_d-1}
      \left(
        \frac{\ifib{d_j}^{m_d}}{2m_d}
        - \frac{1}{2}\log\ifib{d_j}
      \right)
    \Bigg]
  \end{aligned}
\end{equation}

This is the contribution of the fiber system to the overall strain-energy density for this orthotropic symmetry. The elastin fibrils and other non-fibrous constituents composing the matrix are typically modeled as isotropic hyperelastic materials. Here we use a compressible Neo--Hookean ground matrix~\cite{bonetNonlinearSolidMechanics2016}:
\begin{equation}
  \Psi_g =
  \frac{1}{2}\left[
    \frac{E_n}{2(1+\nu)}\left(\trace{C}-3
    -2\log J\right)
    +\frac{\nu E_n}{(1+\nu)(1-2\nu)}(\log J)^2
  \right]
  \label{eq:matrix_energy}
\end{equation}
where \(\bm{C}=\bm{F}^{\mathrm{T}}\bm{F}\), \(J=\det\bm{F}\), and \(\bm{F}\) is the deformation gradient. Additionally, \(E_n\) and \(\nu\) are the small-strain Young's modulus and Poisson's ratio of the Neo-Hookean matrix, respectively. The total strain-energy density for the unit cell is then written as:
\begin{equation}
  \Psi = \phi_f \hat{\Psi}_{\mathrm{fs}} + (1-\phi_f)\Psi_g = \Psi_{\mathrm{fs}} + (1-\phi_f)\Psi_g
  \label{eq:total_energy}
\end{equation}
where \(1 - \phi_f\) is the matrix volume fraction. Overall, the proposed material is a solid mixture of nonlinear, pre-stretched fibers and an isotropic Neo-Hookean matrix with orthotropic symmetry. We analyze and comment on the stability of the proposed material model in~\ref{sec:SI_convexity}.

The complete material model contains 10 parameters: seven for the fiber system, two for the Neo-Hookean ground matrix, and one volume fraction describing their relative contributions. However, one can readily reduce the number of independent parameters by considering special cases. For instance, the material properties describing the axial fibers can be set equal to those describing the diagonal fibers, i.e., \(E_f = E_a = E_d\) and \(m_f = m_a = m_d\). This effectively reduces the number of independent fiber system parameters from seven to five at the cost of reduced flexibility. Similarly, \(\phi_f\) can be set a priori from knowledge of the underlying material system. This assignment has the additional benefit that it effectively removes the implicit length scale dependency from the unit cell size and separates the effect of fiber volume fraction from fiber stiffness during calibration. We use and recommend a preset value of \(\phi_f\) in subsequent applications of the material model.

It is important to note that the aforementioned simplifications do not compromise the anisotropy, which is set by the direction cosines. The direction cosines can be equated to \(\alpha_1 = \alpha_2 = \alpha_3 = 1/\sqrt{3}\) for a cubic anisotropic case where all axial fibers have the same length and the contribution of diagonal fibers is equal along these directions. Development of an isotropic equivalent requires an additional assumption to ensure proper symmetry. Under one possible approach, the equivalent isotropic material model is a summation of isotropic Ogden and Arruda-Boyce hyperelastic model terms with a logarithmic volumetric term. This is presented in~\ref{sec:SI_isotropic_eng}.

\subsection{Stress and Tangent}
The second Piola--Kirchhoff stress contribution from the fiber system is:
\begin{equation}
  \begin{aligned}
    \bm{S}_{\mathrm{fs}} =2\frac{\partial\Psi_{\mathrm{fs}}}{\partial\bm{C}} = \Phi_f \Bigg[
      &\sum_{i=1}^{3}
      \frac{E_a\alpha_i\lambda_o^{2}}{2m_a-1}
      \left(\ifib{a_i}^{m_a-1}-\frac{1}{\ifib{a_i}}\right)
      (\tilde{a}_i\otimes\tilde{a}_i)\\
      &+\sum_{j=1}^{4}
      \frac{E_d\lambda_o^{2}}{2m_d-1}
      \left(\ifib{d_j}^{m_d-1}-\frac{1}{\ifib{d_j}}\right)
      (\tilde{d}_j\otimes\tilde{d}_j)
    \Bigg]
    \label{eq:fiber_stress}
  \end{aligned}
\end{equation}

where \(\Phi_f = \phi_f/(\alpha_1 + \alpha_2 + \alpha_3 + 4)\). The \(\tilde{a}_i\) and \(\tilde{d}_j\) are the unit vectors along the axial and diagonal fiber directions. The total stress is \(\bm{S}=\bm{S}_{\mathrm{fs}}+(1-\phi_f)\bm{S}_g\), where \(\bm{S}_g\) is the Neo-Hookean matrix contribution available in standard texts, e.g.,~\cite{bonetNonlinearSolidMechanics2016}. Similarly, the contribution of the fiber system to the material tangent is written as:
\begin{equation}
  \begin{aligned}
    \mathcal{C}_{\mathrm{fs}} &= 2\frac{\partial \bm{S}_{\mathrm{fs}}}{\partial \bm{C}} =
    \Phi_f\sum_{i=1}^{3}
    \frac{2E_a\alpha_i\lambda_o^{4}}{2m_a-1}
    \left[(m_a-1)\ifib{a_i}^{m_a-2}+\frac{1}{\ifib{a_i}^2}\right]
    (\tilde{a}_i\otimes\tilde{a}_i)
    \otimes(\tilde{a}_i\otimes\tilde{a}_i)\\
    &+\Phi_f\sum_{j=1}^{4}
    \frac{2E_d\lambda_o^{4}}{2m_d-1}
    \left[(m_d-1)\ifib{d_j}^{m_d-2}+\frac{1}{\ifib{d_j}^2}\right]
    (\tilde{d}_j\otimes\tilde{d}_j)
    \otimes(\tilde{d}_j\otimes\tilde{d}_j)
    \label{eq:fiber_tangent}
  \end{aligned}
\end{equation}

\subsection{Material Parameter Calibration}
\label{sec:param_calibration}
Material-parameter calibration for nonlinear anisotropic constitutive models is challenging because standard mechanical tests, including uniaxial and biaxial tests, do not generally admit closed-form stress--stretch solutions. The difficulty is compounded by the high-dimensional and constrained parameter space of the present model. We consider the volume fraction of fibers, \(\phi_f\), to be known, e.g., from experimental results and thus not part of the calibration. The independent material parameters and the constraints are:
\begin{enumerate}
  \item The direction cosines, \(\alpha_1, \alpha_2\), subject to the constraint \(\alpha_1^2 + \alpha_2^2 + \alpha_3^2 = 1\) and individually bounded to the range \(0 < \alpha_i < 1\). A parameterization using spherical coordinates can be adopted to automatically satisfy the constraints.
  \item The fiber stiffness parameters, \(E_a\) and \(E_d\), subject to the constraints \(E_a > 0\) and \(E_d > 0\).
  \item The nonlinear stiffening exponents, \(m_a\) and \(m_d\), subject to the constraints \(m_a > 1\) and \(m_d > 1\).
  \item The ground-matrix material properties, here the Neo-Hookean material parameters, \(E_n\) and \(\nu\), subject to the constraints \(E_n > 0\) and \(0 < \nu < 0.5\).
  \item Optionally, the pre-stretch, \(\lambda_o\), subject to the constraint \(\lambda_o > 0\).
\end{enumerate}

In the literature, inverse FEA is commonly adopted for material parameter calibration, where material parameters are identified by coupling FEA with an optimization algorithm~\cite{chen_finite_2025}. However, derivative-free methods such as Nelder-Mead become computationally expensive when each objective-function evaluation requires a full finite element solve. This limitation is particularly pronounced for the proposed octahedral constitutive model, which involves an eight-dimensional parameter space. To mitigate this computational cost, we streamline parameter calibration by coupling FEBio finite element simulations with an automatic differentiation optimization framework implemented in JAX~\cite{maasFEBioFiniteElements2012a, jax2018github}. This procedure follows the general idea of Finite Element Model Updating (FEMU)~\cite{chen_finite_2025, alhassaniehEfficientMaterialModel2025}. The details of our implementation are presented in~\ref{sec:SI_param_calibration}.

\section{Results}
\subsection{Modeling General Behavior of Fibrous Materials}
\subsubsection{Parameter Calibration to Experimental Data}
\label{sec:calibration_results}

To demonstrate the model's applicability to biomechanical systems, we first applied the approach from Section~\ref{sec:param_calibration} to calibrate the material model to porcine mitral heart valve tissue properties measured by Ross et al.~\cite{rossBiaxialMechanicalCharacterizations2019}. The calibration procedure optimized the material parameters using five displacement-controlled test protocols from the study simultaneously: biaxial; pure shear in the circumferential direction; constrained uniaxial stretching in the circumferential direction; pure shear in the radial direction; and constrained uniaxial stretching in the radial direction. In the pure-shear tests, extension along the loading direction is accompanied by shortening in the orthogonal direction to approximately preserve the specimen area. In the constrained-uniaxial tests, the orthogonal direction remains fixed.

We fixed the fiber volume fraction at a conservative value of \(\phi_f = 0.1\). Since the experimental data are from excised valve tissue, we did not account for in vivo pre-stretch during calibration and set \(\lambda_o = 1\). The optimized values of the material parameters are \(E_a = 18.79\) kPa, \(E_d = 23.91\) kPa, \(m_a = 7.66\), \(m_d = 14.97\), \(E_n = 20.89\) kPa, \(\nu = 0.47\), \(\alpha_1 = 0.92\), and \(\alpha_2 = 0.38\). The calibrated material model response is plotted with the experimental data in Figure~\ref{fig:calibrated_results}. The fit metrics and results of sensitivity analyses are reported in~\ref{sec:SI_param_calibration}.

The test sample is stiffer in the circumferential direction. This is captured by the large \(\alpha_1\) in this setup. Additionally, the small \(\alpha_3 \approx 0.1\) indicates the material behavior in the normal direction is mostly captured by the matrix. The large value of \(\nu\) is likely due to the hydrated nature of the experimental samples and sample data involving volume-preserving shear tests. Overall, the calibrated model captures the material behavior well, consistent with physical expectations. Finally, in \ref{sec:SI_param_calibration}, we also compare our model to an established anisotropic model, namely compressible HGO~\cite{pelleritoComputationalModelingTissue2026}. The proposed model appears to be a better fit for the present dataset than HGO.

\begin{figure}
  \centering
  \includegraphics[width=\linewidth]{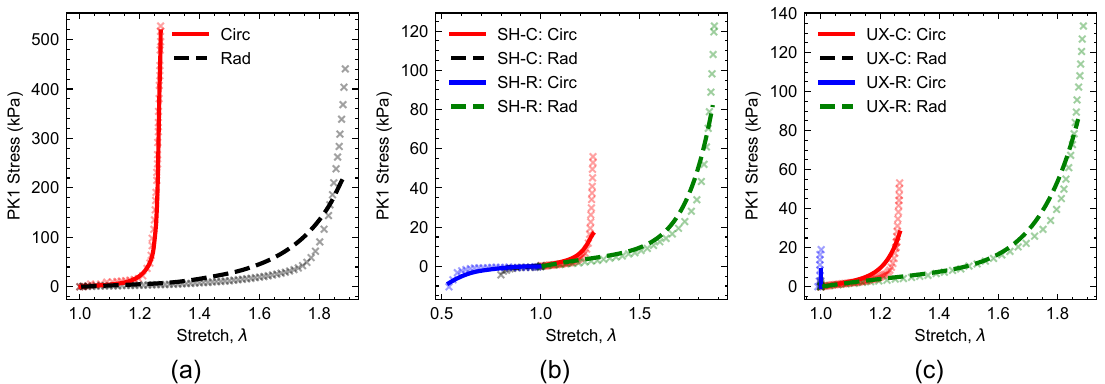}
  \caption{Comparison of the calibrated material model response with experimental data from Ross et al.~\cite{rossBiaxialMechanicalCharacterizations2019}. The fit metrics are reported in Table~\ref{tab:r2-calibration} (\ref{sec:SI_param_calibration}). The biaxial test case is shown in (a). The pure-shear experiments with extension along circumferential (SH-C) and radial (SH-R) directions are shown in (b). The constrained uniaxial test cases are shown in (c) for the same directional setup. In all cases, solid and dashed lines indicate the calibrated material's response in circumferential and radial directions, respectively. The experimental data are shown as points in matching colors.}
  \label{fig:calibrated_results}
\end{figure}

\subsubsection{Modeling Stiffening and Volumetric Behavior}
In the proposed model, the exponents \(m_a\) and \(m_d\) primarily capture the stiffening behavior of the material. The direction cosines, \(\alpha_i\), also modulate the response, as evident from Equation~\ref{eq:fiber_stress}. First, for axial fibers, they act as scaling parameters. Second, they influence the projection of the force generated in the diagonal fibers in the direction of the axial fibers through \(\tilde{d}_j \otimes \tilde{d}_j\), a mechanism analogous to fiber recruitment in the loading direction in fibrous materials. We compare the effect of these parameters in Figure~\ref{fig:stress_volume}a in a simulated uniaxial tension test under displacement control for a few selected cases. Here, the circumferential direction is aligned with the loading axis. We first scaled the exponents obtained from the calibration procedure in Section~\ref{sec:calibration_results} and compared them with the baseline (calibrated model) to demonstrate the effect of the exponents on the material response. Larger exponents produce stronger strain stiffening, whereas smaller exponents produce weaker stiffening. Similarly, swapping the direction cosines, i.e., \(\alpha_1 \leftrightarrow \alpha_2\), changes the directional stiffening response. The swap reorients the diagonal fibers, reducing their projected force along the loading direction, and reduces the axial fiber contribution through its scaling factor. We revisit this change in the mitral valve case study, where it compromises the valve's ability to close properly.

The material model is designed to be highly compressible and can accommodate large volumetric deformations. This feature is particularly important for simulating biological tissues that often experience significant volume reduction under mechanical loading~\cite{ehretInversePoroelasticityFundamental2017,wahlstenCompressibilityPoroelasticityHuman2019,woessnerThreeDimensionalQuantificationCollagen2021,brownMultiscaleMechanicsFibrin2009,banStrongTriaxialCoupling2019}. The volumetric behavior of the material model is controlled by the same parameters as those governing strain stiffening. Specifically, increasing the exponents leads to a steeper volume reduction in both tension and compression compared to the baseline. In Figure~\ref{fig:stress_volume}b, we illustrate this behavior for the selected cases. The general behavior is such that the volume ratio initially increases beyond unity under tension and then decreases as a function of stretch. The stretch range over which the volume ratio exceeds unity is controlled by the specific combination of the parameters describing the fiber system and the matrix. The material model can generate a range of volumetric response curves and can therefore be calibrated against experimental data when available. In other words, if volumetric deformation data is available from experiment, it should be used during calibration procedure to model the material response accurately. Due to the flexibility of the proposed model, we expect that the material model will be able to capture it well. We show such an example in~\ref{sec:SI_case_study_extras}. Finally, the matrix model may be modified to impose strong bulk resistance or incompressibility in the material model.

\begin{figure}
  \centering
  \includegraphics[width=\linewidth]{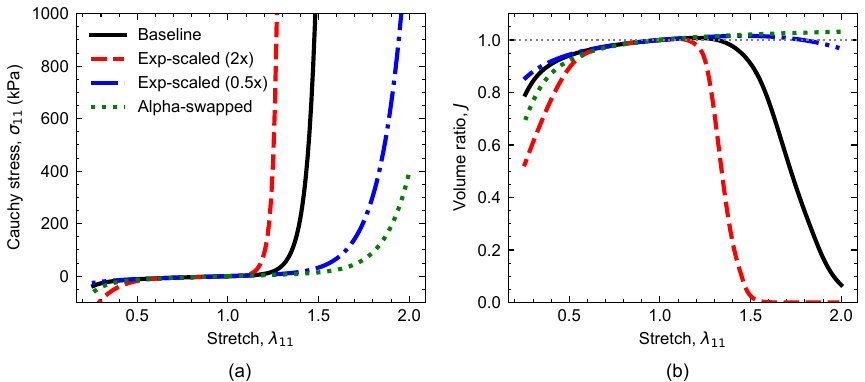}
  \caption{Response of the calibrated material model under uniaxial loading: (a) Cauchy stress and (b) volume ratio. The baseline is compared with cases where the exponents are scaled by 2 or 0.5 and where the direction cosines are swapped (\(\alpha_1 \leftrightarrow \alpha_2\)). The legends in panel (a) apply to both panels.}
  \label{fig:stress_volume}
\end{figure}

\subsubsection{Shear Deformation and Poynting Effect}
Biological soft tissues and their analogs are known to exhibit a strong reverse Poynting effect~\cite{janmeyNegativeNormalStress2007, destradeDominantNegativePoynting2015, destradeCancelingElasticPoynting2023, horganPoyntingEffectsSoft2025}. In this phenomenon, the material tends to contract in the direction transverse to the applied shear. Due to the specific arrangement of the fibers in the unit cell, our material model captures this behavior. We demonstrate this for two cases: (i) constrained simple shear and (ii) rotation of a cylinder that decreases its axial length. Here we present the numerical results for both cases with \(\phi_f = 1\) to isolate the contribution of the fibers to the observed Poynting effect while keeping the other material parameters set to their baseline values from Section~\ref{sec:calibration_results}.

Figure~\ref{fig:poynting_effect}a shows the material response under constrained shear. Here the shear direction is aligned with the circumferential fiber direction in heart valve tissue. Following~\cite{mihaiPositiveNegativePoynting2011}, the Poynting effect is the presence of unequal normal stresses, i.e., \(\sigma_{xx} \neq \sigma_{yy}\) and the reverse Poynting effect occurs when the sheared parallel faces in simple shear tend to contract in the transverse direction, i.e., \(\sigma_{yy} > 0\). The numerical evidence demonstrates that the proposed material model exhibits the reverse Poynting effect. We analytically show that this is the general behavior of the material model in~\ref{sec:SI_reverse_poynting_sshear} which establishes \(\alpha_1^2 < 8/9, m_d > 4\) as a sufficient condition. Finally, the shear stress-strain response in Figure~\ref{fig:poynting_effect}a demonstrates the model's ability to capture a nonlinear shear response.

\begin{figure}
  \centering
  \includegraphics[width=\linewidth]{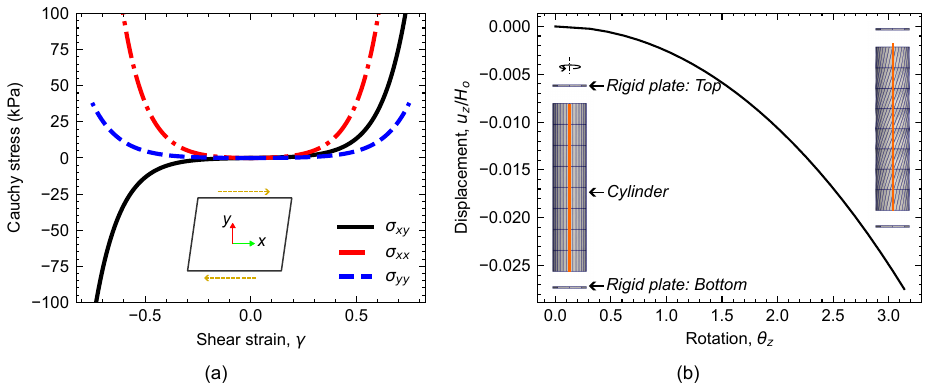}
  \caption{Reverse Poynting effect in two loading cases: (a) normal and shear stresses under constrained simple shear and (b) axial contraction of a cylindrical sample under rotation. In (b), the orange line in the sketch represents the initial length of the cylinder. The axial displacement is normalized by its length, \(H_o=10\,\mathrm{mm}\). }
  \label{fig:poynting_effect}
\end{figure}

In numerical experiments with a cylindrical sample, we applied a rigid-body rotation at the top surface while keeping the bottom surface fixed. We used a cylindrical coordinate system such that the circumferential direction is aligned with the circumferential fibers in heart valve tissue from Section~\ref{sec:param_calibration}. We observed a decrease in the axial length of the cylinder, consistent with the reverse Poynting effect. This is shown in Figure~\ref{fig:poynting_effect}b as a function of the applied rotation.

We note that, in general cases where the matrix is considered, i.e., \(\phi_f < 1\), the reverse Poynting effect may still be observed but could be influenced by the matrix contribution. Similarly, fiber reorientation may alter the observed Poynting effect depending on the new fiber alignment relative to the applied deformation.

\subsection{Case Study: Mitral Valve Deformation}\label{sec:case_study}
Numerous experimental studies have established that heart valve tissue exhibits anisotropic behavior along orthogonal directions~\cite{meadorDetailedMechanicalMicrostructural2020, sadeghiniaBiomechanicsMitralValve2022, phamJBMR2014}. Here, we use our model to simulate an idealized mitral valve with orthotropic material properties. We also investigate the effects of fiber orientation, pre-stretch, and collagen volume fraction on the mechanical behavior of the mitral valve. The results reproduce observations reported in the literature and suggest complications that may emerge due to changes in the microstructural properties of the tissue.

\begin{figure}
  \centering
  \includegraphics[width=0.75\linewidth]{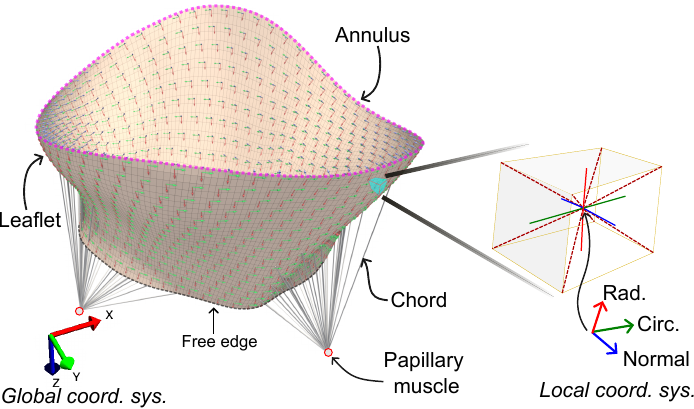}
  \caption{Idealized FE model of the mitral valve showing fiber orientations and the use of the unit-cell-based material model. The primary fiber directions are indicated by the colors green (circumferential), red (radial), and blue (normal).}
  \label{fig:mitral_geom}
\end{figure}

\subsubsection{FEA Model and Fiber Orientation Setup}
For this case study, we used an idealized FE model of the mitral valve shown in Figure~\ref{fig:mitral_geom}. We considered the leaflet material to be orthotropic with fibers oriented along the circumferential and radial directions as shown in the figure. We used the proposed material model only for the leaflets, which were modeled using 4-noded shell elements of thickness 0.39 mm. The in vivo pre-stretch in heart valve leaflets generally differs between the circumferential and radial directions. In the simulations, we assumed an isotropic pre-stretch, \(\lambda_o = 1.3\) following suggestions from Refs.~\cite{rauschMechanicsMitralValve2013a, aggarwalInvivoHeterogeneousFunctional2016, leeInVivoFunctionMitral2017,oomenGrowthRemodelingOpposing2018}. Hence, we note that the idealized FE geometry from Figure~\ref{fig:mitral_geom} represents an unpressurized state, not necessarily an equilibrated configuration due to the pre-stress. In other words, the simulation starts from a pre-stressed configuration rather than a stress-free one. From simulation with boundary constraints but with no external pressure, we find that pre-stretch leads to a mean displacement of approximately 1.5 mm, compared to approximately 9 mm at full pressured state presented in the later part of this section.

Several techniques, e.g., that in Ref.~\cite{rimOptimizationbasedFiberOrientation2015}, can be used to generate the fiber orientation that is compatible with the geometry of the valve and experimental observations. Here, we used a differential-equation-based approach in which the local fiber orientations are defined based on the solution of Poisson's equation as implemented in the FEBio FEA package (Fiber Generator Tool). In this approach, scalar boundary values are first defined at the annular and free edges of the leaflets. The solution of the boundary value problem is then computed, and its gradient gives the first local (fiber) direction. For shell elements, the element normals are considered as the second local (fiber) direction. Finally, their cross-product completes the set. The fiber orientation thus generated, and visualized in Figure~\ref{fig:mitral_geom}, is visually comparable to those reported based on experimental techniques~\cite{sadeghiniaBiomechanicsMitralValve2022, meadorDetailedMechanicalMicrostructural2020, rossBiaxialMechanicalCharacterizations2019}. In our FEA, the circumferential direction (green) is associated with \(\alpha_1\), radial (red) with \(\alpha_2\), and normal (blue) with \(\alpha_3\). The fiber orientations can be controlled by prescribing scalar values in additional locations on the leaflet or through manual perturbation of the generated vectors.

\begin{figure}
  \centering
  \includegraphics[width=\linewidth]{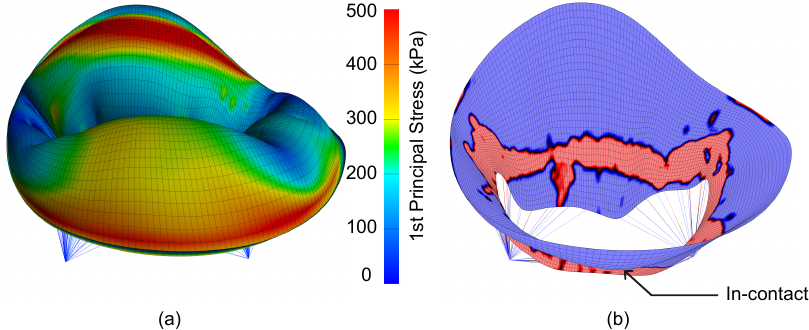}
  \caption{(a) Deformed configuration of the mitral valve at maximum applied pressure (10 kPa). (b) The contact areas between the leaflets at the maximum pressure (in red, shown in reference configuration).}
  \label{fig:mitral_closure}
\end{figure}

The chordae tendineae were modeled using line elements with one end affixed to random locations on the leaflet and the other end terminating at one of two fixed points that represent the papillary muscles. In the context of FEA, we utilized elastic truss elements for this purpose with an Ogden material model defined below~\cite{ogdenNonlinearElasticDeformations1997,maasFEBioFiniteElements2012a}:

\begin{equation}\label{eq:ogden_mat}
  \Psi_{\mathrm{c}} = \frac{1}{2}c_p(J - 1)^2 + \sum_{i=1}^N \frac{c_i}{m_i^2}(\lambda_1^{m_i}+\lambda_2^{m_i}+\lambda_3^{m_i}- 3 - m_i \log J)
\end{equation}

The choice of this material model is motivated by the large exponents for the leaflet material after calibration, similarity to the isotropic equivalent of the proposed model, and its widespread use in related literature. Alternatives, such as linear truss elements, are also possible, but they typically lead to stress concentrations at the chordal attachment points on the leaflet in our experience -- likely due to a stiffness mismatch between the leaflet and chordae, coupled with the simplified line representation of the chordae. In this work, we chose \(N = 1\), \(c_p = 1000\,\mathrm{kPa}\), \(c_1 = 100\,\mathrm{kPa}\), and \(m_1 = 8\) to be the parameter values for the chordae.

We applied a physiologically relevant pressure of 10 kPa to the outer surface of the leaflet. Additionally, a normal displacement was applied to the elements adjacent to the annular ring as a first-order approximation of the motion of the valve in vivo. We used potential-based contact mechanics in the simulation for the leaflet with parameter values \(k_c = 1\), \(p = 4\), \(r_\mathrm{in} = 0.2\,\mathrm{mm}\), and \(r_\mathrm{out} = 0.5\,\mathrm{mm}\) that describe the strength of the contact interaction as a function of the contact gap~\cite{kamenskyContactFormulationBased2018}.

\subsubsection{Deformation of Mitral Valve and Role of Fiber Orientation}
The deformed configuration of the mitral valve for the fitted material parameters at 10 kPa pressure is shown in Figure~\ref{fig:mitral_closure}a. Closure occurred at a lower pressure, approximately 8 kPa. The self-contact-derived coaptation region in the leaflet is illustrated in Figure~\ref{fig:mitral_closure}b, further confirming the closure of the valve.

The distribution of first principal stress in the leaflets is shown using the color map in Figure~\ref{fig:mitral_closure}a. The overall stress pattern indicates a right-skewed log-normal-like distribution. We measured a mean and median stress of 194.36 and 145.75 kPa, respectively. The fraction of surface area in contact is approximately 0.25 in the closed configuration.

\subsubsection{Effect of Fiber Reorientation}
We now numerically investigate the effects of fiber reorientation on the mechanical response of the mitral valve leaflets. Specifically, we first examine how changes in the local fiber orientation affect the stress distribution and overall deformation of the leaflets under physiological loading conditions. We then analyze the impact of a widespread fiber reorientation on the deformation.

The local fiber reorientation was generated through additional prescribed scalar values for a few selected elements in the anterior and posterior belly region of the leaflet. The resulting fiber orientation is visualized in Figure~\ref{fig:local_reorient}a. This represents a localized heterogeneity in the fiber orientation compared to the baseline. The corresponding deformed configuration is shown in Figure~\ref{fig:local_reorient}b. Here, the color indicates the value of the first principal stress. Compared with the baseline configuration in Figure~\ref{fig:mitral_closure}a, the localized perturbation did not compromise valve competence. However, the stress distribution changed locally in the regions with altered fiber orientation. This is quantitatively shown in Figure~\ref{fig:local_reorient}c where we calculate the relative difference as \(\Delta \sigma_1/\sigma_{1, \mathrm{ref}}\), where \(\sigma_1\) is the first principal stress and \(\Delta \sigma_1\) is the elementwise difference between this case and the reference case. The stress values from Figure~\ref{fig:mitral_closure}a are used as the reference in this measure. This view highlights the sensitivity of the stress distribution to local changes in fiber orientation. Notably, such sensitivity is only apparent in stress- or strain-related quantities, as we did not observe any pronounced persistent signature in the displacement fields (see~\ref{sec:SI_case_study_extras}). We suspect the effects are masked in the displacement field by the large rotation of the leaflet elements.

\begin{figure}
  \centering
  \includegraphics[width=\linewidth]{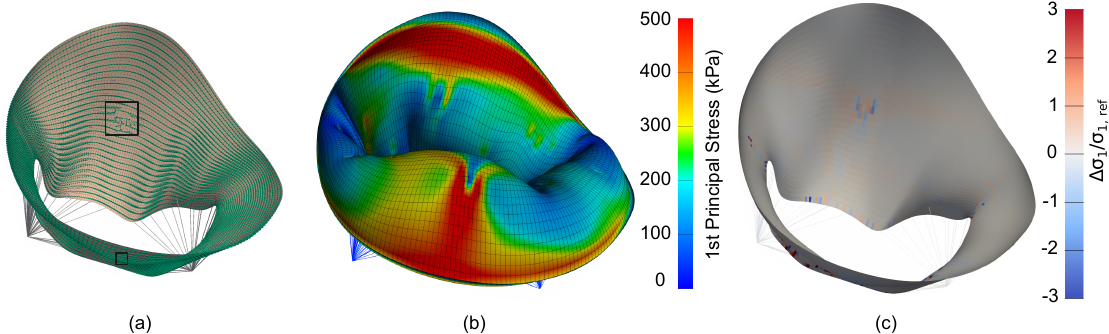}
  \caption{Effect of localized fiber reorientation on the mitral valve leaflets. (a) The modified fiber orientations with localized perturbations in the belly regions compared to the baseline (marked by black squares). (b) The deformed configuration of the valve at a maximum applied pressure of 10 kPa showing proper closure. (c) The relative difference in the first principal stress compared to the baseline configuration, i.e., elementwise relative difference between (b) and baseline (Figure~\ref{fig:mitral_closure}a).}
  \label{fig:local_reorient}
\end{figure}

A clear effect in the displacement field is, however, visible in cases where there is a widespread fiber reorientation throughout the leaflet. In particular, we consider a case where the radial direction of the material is stiffer than the circumferential direction. This kind of reversed response has been observed experimentally with possible links to calcium deposition in the leaflet~\cite{sadeghiniaBiomechanicsMitralValve2022, phamJBMR2014}. Here, we modified the material behavior by swapping the values of \(\alpha_1\) and \(\alpha_2\) for all elements in the leaflet. From the fiber orientation perspective, this represents a rotation of the fiber directions by 90 degrees about the surface normal. The deformed configuration at the maximum pressure of 10 kPa is shown in Figure~\ref{fig:global_reorient_sv}a where a gap remains between the leaflets. This result illustrates the sensitivity of simulated coaptation to leaflet anisotropy, a factor also examined in prior work~\cite{kunzelmanFluidStructureInteraction2007, vandemaeleAlterationsHumanMitral2022}.

The sliced views for the three cases considered so far are compared in Figure~\ref{fig:global_reorient_sv}b, summarizing these observations. The section view shows a residual gap after global fiber reorientation. The gap persisted in a subsequent simulation at higher pressure. Comparing the case with localized reorientation to the baseline, we observed no pronounced effect in the anterior region but some effect in the posterior region where fiber reorientation was introduced. However, these displacement differences varied during the simulations and were generally more pronounced in the folding regions than at the reorientation sites (see~\ref{sec:SI_case_study_extras}).

\begin{figure}
  \centering
  \includegraphics[width=\linewidth]{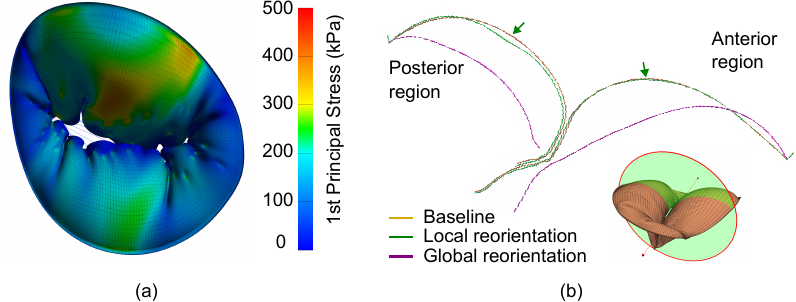}
  \caption{(a) Deformed configuration of the mitral valve with a residual gap between the leaflets after global fiber reorientation. (b) Sliced view of the deformed configuration of the mitral valve for various cases considered in this study. The regions of localized fiber reorientation are marked by arrows.}
  \label{fig:global_reorient_sv}
\end{figure}

\subsubsection{Effect of Collagen Volume Fraction and Pre-Stretch}
In this section, we examine the effects of collagen volume fraction, \(\phi_f\), and isotropic pre-stretch, \(\lambda_o\), on the deformation of the heart valve. Specifically, we present a phase diagram describing their effect on the deformation of the valve. To this end, we started with the baseline model and material parameters from Figure~\ref{fig:mitral_closure}, with \(\phi_f = 0.1\) and \(\lambda_o = 1.3\). We then systematically varied these parameters in increments of 0.10 to investigate their individual and combined effects while keeping other material parameters unchanged.

From the strain energy defined in Equation~\ref{eq:fiber_system_energy}, the effect of \(\phi_f\) is to modulate the contribution of the fiber system to the overall energy function. Especially for large exponent values, increasing \(\phi_f\) stiffens the unit cell in tension because of the increased contribution of the fiber system and has the opposite effect in compression due to the reduced matrix contribution in the small to moderate strain range. Pre-stretch, on the other hand, controls the effective portion of the constitutive relation utilized by the model. In fact, according to Equations~\ref{eq:fiber_stress} and~\ref{eq:fiber_tangent}, the material stress and stiffness scale quadratically and quartically, respectively, with pre-stretch directly, with additional effects through the invariants. Thus, under fixed load, it reduces the maximum displacement/strain relative to the at-rest configuration under a given pressure, which can reduce leaflet coaptation in the model.

We present the phase diagram in Figure~\ref{fig:phase_diagram}a. From bottom to top in this figure, the pre-stretch, \(\lambda_o\), increases from 1.0 (no pre-stretch) to 1.4, and from left to right the collagen volume fraction, \(\phi_f\), increases from 0.1 to 0.4. The numerical values in the cells represent the mean value of the first principal strain in the leaflet (measured with respect to an unpressurized configuration) and the fraction of leaflet area in contact (in parentheses). All the reported numerical values are normalized by the corresponding results from the baseline case.

The results indicate that increasing pre-stretch reduces the contact area. Reducing pre-stretch, on the other hand, increases contact area but leads to billowing in the leaflet. For instance, at a fixed \(\phi_f = 0.1\), changing \(\lambda_o\) to 1 and 1.4 led to a 67\% increase and a 16\% decrease in the contact area compared to the baseline, respectively. Billowing is further evidenced by the increasing strain in the leaflet and the deformed configuration shown in the figure with no pre-stretch. Decreased pre-stretch is also associated with increased strain which is clearly visible in the leaflet segments near the free edge. Without pre-stretch, the mean strain was more than twice the baseline value. The range of first principal strain (minimum, maximum; measured with respect to the unpressurized state) also increased substantially from (0.04, 1.03) at baseline to (0.12, 2.22) with no pre-stretch. Both mean principal strain and contact area decreased with increasing pre-stretch, but mean principal strain decreased more rapidly (Figure~\ref{fig:phase_diagram}b). Finally, the reduction in contact area was associated with a reduction in contact pressure. The reduction in measured strain, however, was not accompanied by a reduction in principal stress as the stress incorporates the effect of pre-stretch in fibers as well. For the \(\phi_f = 0.1\) case, for example, the mean stress showed a nonlinear increase with the pre-stretch starting beyond \(\lambda_o = 1.2\). These trends are visualized in Figure~\ref{fig:phase_diagram}b.

Increasing the collagen volume fraction, in general, decreased the contact area and average strain. This effect is consistent with the expected stiffening of the leaflet due to higher fiber content. However, the effect is not as pronounced as that of pre-stretch, over the range of collagen volume fractions examined here. The contact area, for example, decreased from 1.67 to 1.54 as \(\phi_f\) increased from 0.1 to 0.4 at \(\lambda_o = 1.0\). At \(\phi_f =0.1\), increasing \(\lambda_o\) from 1.0 to 1.10 produced a larger reduction (1.67 to 1.46). Consequently, the increased volume fraction did not notably alter the deformed configuration. Finally, in contrast to pre-stretch, increasing \(\phi_f\) consistently decreased the mean strain, stress, contact area, and contact pressure, as shown in Figure~\ref{fig:phase_diagram}c for \(\lambda_o = 1.0\).

\begin{figure}
  \centering
  \includegraphics[width=\linewidth]{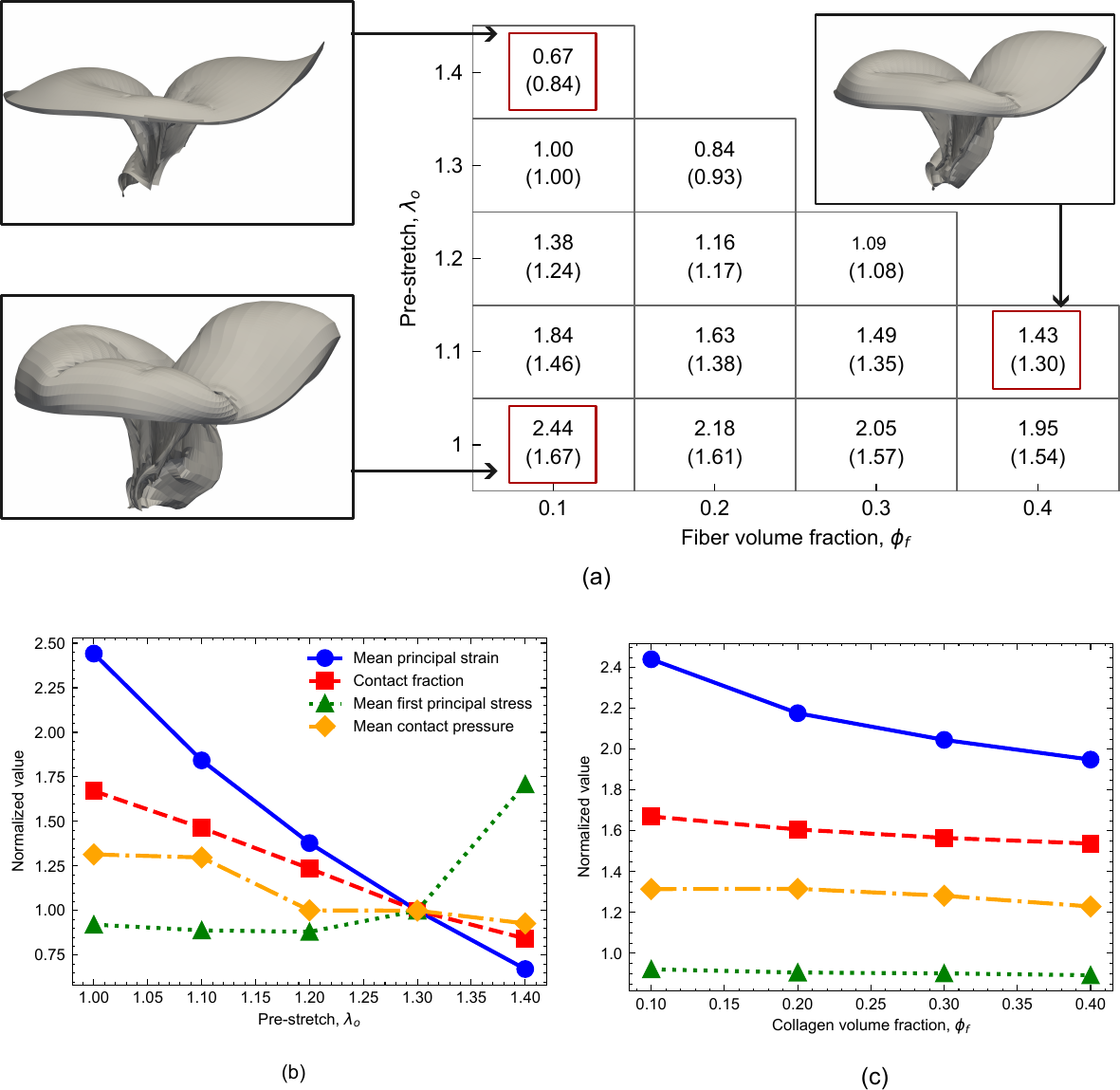}
  \caption{(a) Phase map showing the effect of collagen volume fraction, \(\phi_f\), and pre-stretch in the fibers, \(\lambda_o\). Each cell shows the normalized mean first principal strain, with the normalized fraction of leaflet area in contact in parentheses. The normalization case is marked as the baseline and corresponds to \((\lambda_o, \phi_f) = (1.3, 0.10)\). Deformed configurations are shown for selected cases. (b) The change in normalized values for mean first principal stress and strain, fraction of area in contact, and mean contact pressure with respect to \(\lambda_o\) for \(\phi_f = 0.10\). (c) Same as (b) but plotted along \(\phi_f\) for \(\lambda_o = 1.0\). The legend from panel (b) applies to both (b) and (c).}
  \label{fig:phase_diagram}
\end{figure}

\section{Discussion}
The preceding results demonstrate the ability of the proposed anisotropic hyperelastic model to represent the mechanical behavior of soft fibrous materials. A case study involving mitral valve deformation further supports this claim. The highlights of this work and their implications are briefly discussed in this section.

\subsection{Feasibility of Proposed Constitutive Model}
The primary contribution of this work is the proposed octahedral fibrous material model, which captures several characteristic behaviors of soft tissues. In particular, we demonstrated that the model approximates the exponential stiffening observed in soft biological tissue, with flexibility provided by the material parameters. The model also captures strong volume reduction and the reverse Poynting effect observed in soft materials, the latter due to the particular arrangement of fibers in a cuboid unit cell. Overall, the model is suitable for fibrous materials with orthotropic symmetry, as found in various biological tissues.

\subsection{Effect of Fiber Orientation on Mitral Valve}
In the idealized mitral valve simulations (Section~\ref{sec:case_study}), prescribed changes in fiber orientation altered leaflet stress and strain fields. Swapping the preferred material directions throughout the leaflets also impaired coaptation.

We draw particular attention to the case of localized fiber reorientation. The baseline and locally reoriented models showed similar overall deformation and closure, with differences in the spatial distribution of leaflet stress. Note that principal stresses are associated with directions -- thus direct comparison in Figure~\ref{fig:local_reorient}c is somewhat ambiguous. Nevertheless, the correlation between the difference in the stress field and fiber reorientation suggests that the effect of local perturbation is not fully masked by the surrounding tissue.

This local stress redistribution has the potential to recruit or further orient randomly oriented fibers and subsequently lead to loss of the preferred anisotropy, which is pathological~\cite{kunzelmanFluidStructureInteraction2007}. Therefore, this effect of fiber reorientation warrants further studies to assess the generalizability of the observation, the role of material/model selection, and its long-term effects.

Finally, our analyses suggest it may be easier to detect such maladaptation of the fibers using strain rather than displacement (see~\ref{sec:SI_case_study_extras}). Since strain states can be obtained readily from clinically available data~\cite{wu_geometric_2026}, these results support further investigation into the clinical use of strain information.

\subsection{Effect of Pre-Stretch and Collagen Volume Fraction}
In numerical experiments with fixed geometry and loading, increasing pre-stretch and/or collagen volume fraction generally decreased mean leaflet strain and coaptation area. In the parameter sweep, lower pre-stretch was associated with greater billowing, and higher pre-stretch reduced the coaptation area. The tested range includes the value of approximately 1.3 suggested by prior studies~\cite{rauschMechanicsMitralValve2013a, aggarwalInvivoHeterogeneousFunctional2016, leeInVivoFunctionMitral2017,oomenGrowthRemodelingOpposing2018}.

The decrease in the strain associated with increasing pre-stretch is expected because pre-stretch shifts the fiber response to larger effective stretches. We reiterate, however, that the measured strain is for the shell elements with respect to the unpressurized configuration. The actual strain in the fibers when incorporating the pre-stretch is significantly different. For instance, we recorded effective fiber stretches (minimum, maximum, mean) of (0.72, 1.37, 1.0) in the circumferential direction and (1.15, 2.02, 1.47) in the radial direction.

Collagen fraction had a smaller effect than pre-stretch. Our observations remain broadly consistent with prior work in this area. For instance, Kunzelman et al.~\cite{kunzelmanFluidStructureInteraction2007} reported that an increased collagen volume fraction reduced the regurgitant volume. In the present simulations, increasing collagen volume fraction reduced contact area and had little effect on the deformed configuration at low pre-stretch.

Finally, we emphasize that the results here are primarily focused on the calibrated fiber orientation. Material anisotropy strongly influences valve deformation, as shown in this study and published work. We explored a special case where the material is assumed to be cubic anisotropic, i.e., \(\alpha_1 = \alpha_2 = \alpha_3 = 1/\sqrt{3}\). We present the result in Figure~\ref{fig:FA4_extra_phase_diag_ciso} (in~\ref{sec:SI_case_study_extras}) with a broader range of parameters. In this cubic case, the leaflets showed reduced contact at high pre-stretch across the collagen volume fractions examined. We aim to explore such effects of fiber orientation in a future study.

\subsection{Limitations}
The proposed material model attempts to capture material behavior originating from non-affine deformation of fibers using an affine approximation, which is typical of hyperelastic material models proposed in literature and used in practice. This hyperelastic formulation describes the elastic response of the leaflet tissue and omits rate dependence, loading--unloading hysteresis, stress relaxation, and history-dependent effects such as Mullins-type softening and permanent set. Consequently, the present simulations do not quantify energy dissipation within the leaflet tissue or changes in its mechanical response with repeated loading. Rate effects have been measured in uniaxial and biaxial tests of semilunar valve tissue~\cite{anssariBenamAorticValveViscoelasticity2011,anssariBenamRateDependentSemilunarValves2020}. Biaxial studies of porcine mitral leaflets have reported protocol-dependent rate sensitivity, measurable hysteresis, and stress relaxation~\cite{grashowBiaxialStressStretch2006,grashowBiaxialCreepStressRelaxation2006,jettAnisotropicAtrioventricularValves2018}. These are important physics that should be considered for more accurate prediction of soft material deformation. The proposed hyperelastic model can, however, be easily be incorporated into poro-visco-elasticity as implemented in \cite{yifei}. Isotropic model similar to the one proposed here, for instance, is implemented in poro-elasticity in \cite{konstantinos} and with damage evolution in \cite{angelos} (to give Mullins type softening). Here we discuss limitations related to our material model beyond these assumptions and then those related to the case study.

The primary limitations of the material model are the complexity of the energy function, its relatively large number of parameters, and its strong nonlinearity in power-law form. These features limited our ability to obtain analytical solutions for boundary value problems that are not only useful for material parameter calibration but also shed light on the various deformation mechanisms. We substantially reduced the calibration difficulty with the framework described in Section~\ref{sec:param_calibration}.

Using this framework, we calibrated the material model to five sets of experimental data simultaneously. We obtained an acceptable fit even with tight bounds on the material parameters to reduce difficulties in nonlinear FEA. For instance, a better fit can be achieved by allowing larger exponents. Large exponents in complex material models are not uncommon, e.g.,~\cite{houAutomatedDatadrivenDiscovery2024} and related works cited therein. In our experience, however, larger exponents led to convergence issues in subsequent complex simulations, e.g., heart valve deformation, likely due to \(J\rightarrow 0\) in such cases as shown in Figure~\ref{fig:stress_volume}b. Additionally, the calibrated parameters are more likely to correspond to local minima than to a global minimum, as is the case in numerical optimization more generally. However, the optimizer appeared well constrained due to the incorporation of different mechanical test schemes. For a fixed computing budget, we found the calibrated results to be most sensitive to the choice of initial values for the exponents. When calibration was restricted to the biaxial results, the calibrated parameters did not yield comparable accuracy in the other tests, likely due to the presence of local minima. We have elaborated on these considerations in~\ref{sec:SI_param_calibration}. Despite these limitations, the calibration demonstrates that the optimization approach can efficiently identify material parameters and that the proposed material model is sufficiently flexible for complex materials.

Another limitation of the material model is a lack of unconditional stability due to fiber under compression. Although this did not lead to convergence issues for the boundary value problems we investigate here, it may be limiting for other scenarios. We discuss this at length in~\ref{sec:SI_convexity}. The observations made in the present work related to heart valve deformation also require further assessment, especially against clinical observations. In particular, we draw attention to the case where global fiber reorientation leaves a gap between the simulated leaflets. With a more realistic representation of chordae tendineae, the degree of billowing may differ from the current results. The role of the chordae tendineae has been explored in the literature, e.g.,~\cite{mangineEffectParametricVariation2024}. For completeness, we present one example scenario where softer chordae lead to billowing in~\ref{sec:SI_case_study_extras} and the results are consistent with our physical expectation. In our experience, simplification in the chordae model does not appear to compromise the general applicability of the proposed material model, but the case-study observations warrant further investigation into the role of microstructural properties in biomechanical systems.

\section{Conclusion}
In this work, we developed an anisotropic hyperelastic material with orthotropic symmetry to model soft fibrous materials. The material model contains ten free parameters that can be calibrated to capture the mechanical behavior of various soft tissues, such as strain stiffening and the reverse Poynting effect. We provide guidance toward efficient material parameter calibration and demonstrate its applicability to soft-tissue mechanics through a case study of mitral valve deformation. The case study illustrates how prescribed changes in leaflet anisotropy can affect stress distributions and coaptation in an example simulation of a mitral valve. The material model is equally applicable to general use cases such as modeling tissue grafts or fiber-reinforced composites. Due to the flexibility of the proposed model, observed stability for complex boundary value problems, and its capability to capture various key behaviors of soft tissue, we expect it to be valuable to a broader community of scientists and engineers.

\section*{Data and Code Availability}
Relevant data and computer code are available at \href{https://github.com/wu-lab-crisp/octahedral-fibrous-material-model}{Octahedral Fibrous Model.}

\section*{Acknowledgment}
This work was supported by an Additional Ventures Single Ventricle Research Fund award and the National Institutes of Health NHLBI K25 HL168235 award. PKP acknowledges a seed grant from University of Pennsylvania Materials Research Science and Engineering Center (MRSEC) NSF DMR-2309043. We thank Dr. Steve Maas of the University of Utah for helpful discussions related to the FEBio plugin for the material model.

\section*{CRediT authorship contribution statement}
\textbf{Nishan Parvez:} Conceptualization, Investigation, Methodology, Software, Validation, Formal analysis, Visualization, Writing -- Original Draft.
\textbf{Prashant K. Purohit:} Conceptualization, Methodology, Resources, Supervision, Writing -- Review \& Editing.
\textbf{Wensi Wu:} Conceptualization, Funding acquisition, Methodology, Resources, Supervision, Writing -- Review \& Editing.

\section*{Declaration of AI and AI-assisted technologies in the writing process} During the preparation of the work, the authors used generative AI tools to (i) perform routine work related to data cleaning, organization, and formatting, (ii) assist in drafting the manuscript through language suggestions, grammatical correction, and reference management, and (iii) perform preliminary review to assess the overall quality and coherence of the manuscript. After using these tools, the authors reviewed and edited the content as needed and take full responsibility for the content of the publication.

\appendix
\counterwithin{figure}{section}
\counterwithin{table}{section}
\renewcommand{\thefigure}{\Alph{section}.\arabic{figure}}
\renewcommand{\thetable}{\Alph{section}.\arabic{table}}

\section{Fiber Directions}
\label{sec:SI_fiber_dirs}
Let \(\tilde{a}_i\) (\(i=1,2,3\)) be an orthonormal triad of local material axes and let \(L_d\) denote the common reference length of the diagonal fibers. The axial fiber vectors are \(\vec{a}_i=L_{a_i} \tilde{a}_i = L_d\alpha_i\tilde{a}_i\). The four unit diagonal directions are:
\begin{equation}
  \begin{aligned}
    \tilde{d}_1 &= \alpha_1\tilde{a}_1+\alpha_2\tilde{a}_2+\alpha_3\tilde{a}_3 &
    \tilde{d}_2 &= \alpha_1\tilde{a}_1-\alpha_2\tilde{a}_2+\alpha_3\tilde{a}_3\\
    \tilde{d}_3 &= \alpha_1\tilde{a}_1+\alpha_2\tilde{a}_2-\alpha_3\tilde{a}_3 &
    \tilde{d}_4 &= \alpha_1\tilde{a}_1-\alpha_2\tilde{a}_2-\alpha_3\tilde{a}_3
  \end{aligned}
  \label{eq:fiber_directions}
\end{equation}
The squared fiber stretches are the invariants:
\begin{equation}
  I_{a_i}=\bm{C}:(\tilde{a}_i\otimes\tilde{a}_i)
  \qquad
  I_{d_j}=\bm{C}:(\tilde{d}_j\otimes\tilde{d}_j)
  \label{eq:fiber_invariants}
\end{equation}
For a general unit fiber direction \(\tilde{v}_f\), \(\partial I_f/\partial\bm{C}=\tilde{v}_f\otimes\tilde{v}_f\) and \(\partial^2I_f/\partial\bm{C}^2=0\). With the pre-stretch relation from Equation~\ref{eq:prestretch_invariants}:
\begin{equation}
  \frac{\partial\ifib{a_i}}{\partial\bm{C}}=\lambda_o^{2}(\tilde{a}_i\otimes\tilde{a}_i) \qquad
  \frac{\partial\ifib{d_j}}{\partial\bm{C} }=\lambda_o^{2}(\tilde{d}_j\otimes\tilde{d}_j)
\end{equation}

\section{Convexity and Stability of the Strain-Energy Function}
\label{sec:SI_convexity}
The requirement of objectivity for the proposed model is satisfied by construction. It can also be shown that the fiber-system strain-energy density is a convex function of \(\bm{C}\) when the exponents, \(m_a, m_d\), are greater than 1, leading to a positive definite material tangent. For an arbitrary unit fiber direction \(\tilde{v}_f\):
\begin{equation}
  I_f=\bm{C}:(\tilde{v}_f\otimes\tilde{v}_f) = \bm{C}:\bm{V} > 0
  \qquad
  \ifib{f}=\lambda_o^{2}I_f
  \label{eq:convexity_invariant}
\end{equation}
where \(\bm{V} = \tilde{v}_f\otimes\tilde{v}_f\). The individual-fiber energy entering Equation~\ref{eq:fiber_system_energy} has the form
\begin{equation}
  \Psi_f
  =\kappa_f
  \left(
    \frac{\ifib{f}^{m_f}}{m_f}
    -\log\ifib{f}
  \right)
  \label{eq:convexity_fiber_energy}
\end{equation}
where \(\kappa_f>0\) collects the corresponding constants, elastic modulus, and geometric factors. Its first derivative with respect to \(\ifib{f}\) is \(\frac{\mathrm{d}\Psi_f}{\mathrm{d}\ifib{f}} = \kappa_f (\ifib{f}^{m_f-1} - \frac{1}{\ifib{f}})\). The second derivative is:
\begin{equation}
  \frac{\mathrm{d}^2\Psi_f}{\mathrm{d}\ifib{f}^2}
  =\kappa_f\left[(m_f-1)\ifib{f}^{m_f-2}+\frac{1}{\ifib{f}^2}\right]
  \label{eq:convexity_scalar_second_derivative}
\end{equation}
For \(\ifib{f}>0\), this expression is strictly positive when \(m_f \geq 1\). Since \(\partial\ifib{f}/\partial\bm{C}=\lambda_o^2\bm{V}\) and \(\partial^2\ifib{f}/\partial\bm{C}^2=0\), the first derivative with respect to \(\bm{C}\) is:
\[
  \frac{\partial\Psi_f}{\partial\bm{C}}
  =\frac{\mathrm{d}\Psi_f}{\mathrm{d}\ifib{f}}\frac{\partial\ifib{f}}{\partial\bm{C}}
  =\lambda_o^2\frac{\mathrm{d}\Psi_f}{\mathrm{d}\ifib{f}}\bm{V}
\]
The associated Hessian is:
\begin{equation}
  \begin{aligned}
    \frac{\partial^2\Psi_f}{\partial\bm{C}^2}
    &=\frac{\mathrm{d}^2\Psi_f}{\mathrm{d}\ifib{f}^2}
    \frac{\partial\ifib{f}}{\partial\bm{C}}\otimes\frac{\partial\ifib{f}}{\partial\bm{C}}
    +\frac{\mathrm{d}\Psi_f}{\mathrm{d}\ifib{f}}\frac{\partial^2\ifib{f}}{\partial\bm{C}^2}\\
    &=\lambda_o^4\frac{\mathrm{d}^2\Psi_f}{\mathrm{d}\ifib{f}^2}\bm{V}\otimes\bm{V}
  \end{aligned}
  \label{eq:convexity_fiber_hessian}
\end{equation}
which is positive semidefinite with \(m_f \geq 1\). The fiber-system energy, being a positive weighted sum of these terms, is therefore convex in \(\bm{C}\) and the overall Hessian is positive definite when the exponents are all greater than 1. This, however, does not imply positive definiteness of the total tangent. In particular, the geometric part of the tangent may lead to a loss of rank-one convexity as the fibers are severely compressed. Noting \(\ifib{f} = \lambda_o^2 (\bm{F}\tilde{v}_f)\cdot(\bm{F}\tilde{v}_f)\), \(D \ifib{f} (\bm{F})[\bm{H}] = 2 \lambda_o^2 \vec{f}\cdot \vec{h}\) with \(\vec{f} = \bm{F} \tilde{v}_f\) and \(\vec{h} = \bm{H} \tilde{v}_f\). Similarly, \(D^2 \ifib{f} (\bm{F}) [\bm{H}, \bm{H}] = 2 \lambda_o^2 \vec{h}\cdot \vec{h}\).

The first variation of the energy with respect to \(\bm{F}\) is:
\begin{equation}
  D\Psi_f(\bm{F})[\bm{H}] = \frac{\mathrm{d}\Psi_f}{\mathrm{d}\ifib{f}} D\ifib{f}(\bm{F})[\bm{H}] = 2
  \lambda_o^2\frac{\mathrm{d}\Psi_f}{\mathrm{d}\ifib{f}} \vec{f} \cdot \vec{h}
\end{equation}

The second variation of the energy with respect to \(\bm{F}\) is:
\begin{equation}
  D^2 \Psi_f(\bm{F})[\bm{H}, \bm{H}] = \frac{\mathrm{d}^2\Psi_f}{\mathrm{d}\ifib{f}^2} (D\ifib{f}(\bm{F})[\bm{H}])^2 + \frac{\mathrm{d}\Psi_f}{\mathrm{d}\ifib{f}} D^2 \ifib{f}(\bm{F})[\bm{H}, \bm{H}]
\end{equation}

Taking \(\bm{H} = \vec{x}_o \otimes \vec{y}_o\), \(\vec{h} = (\vec{x}_o \otimes \vec{y}_o) \tilde{v}_f= (\tilde{v}_f\cdot \vec{y}_o) \vec{x}_o\), the second variation becomes:
\begin{equation}
  % \begin{aligned}
  D^2 \Psi_f(\bm{F})[\bm{H}, \bm{H}] = 2(\tilde{v}_f\cdot \vec{y}_o)^2 \lambda_o^2 \left[2\lambda_o^2  \frac{\mathrm{d}^2\Psi_f}{\mathrm{d}\ifib{f}^2} (\vec{f} \cdot \vec{x}_o)^2 + \frac{\mathrm{d}\Psi_f}{\mathrm{d}\ifib{f}} \vec{x}_o\cdot \vec{x}_o\right]
  %  &=2(\tilde{v}_f\cdot \vec{y}_o)^2 \lambda_o^2 \vec{x}_o \left[2\lambda_o^2\frac{\mathrm{d}^2\Psi_f}{\mathrm{d}\ifib{f}^2} (\vec{f}\otimes \vec{f}) + \frac{\mathrm{d}\Psi_f}{\mathrm{d}\ifib{f}} \bm{I} \right]\vec{x}_o
  % \end{aligned}
\end{equation}
The rank-one convexity condition requires the second variation to be non-negative. In this particular case, non-negative first and second derivatives provide a sufficient criterion. Although the second derivative satisfies this criterion for any \(\ifib{f}\) with appropriate values of the material parameters, the first derivative (i.e., stress-like quantity related to the geometric stiffness) is negative for \(\ifib{f} < 1\). Thus, a single fiber may lose rank-one convexity under compression. For instance, if the perturbation is such that \(\vec{x}_o \perp \vec{f}\), the rank-one convexity is lost once \(\ifib{f} < 1\). More critically, \(\frac{\mathrm{d}\Psi_f}{\mathrm{d}\ifib{f}} \rightarrow -\infty\) as \(\ifib{f} \rightarrow 0\). Thus, the model may lose rank-one convexity and, as a consequence, polyconvexity, in \(\bm{F}\) at the compressive limit of a fiber. A similar perturbation leads to loss of rank-one convexity in the fiber energy described by a Fung-type formulation, e.g., anisotropic fiber energy contribution in~\cite{wuAnisotropicConstitutiveModel2018}, under compression, although its first derivative remains finite.

At the fiber level, it is possible to circumvent this issue with tension-only fibers~\cite{markertGeneralPolyconvexStrainenergy2005}. However, at unit cell level, this approach leads to instability under general deformation without a ground matrix even in simple loading scenarios. For instance, in a uniaxial tension test, the absence of compressive resistance in transverse fibers results in a collapse of the unit cell due to the presence of coupling diagonal fibers with no resistive capacity in the transverse directions. Thus, the compressive resistance in the fibers is a necessity for the model, even when it leads to complications in certain cases. Alternatives, such as incorporating unit cell level properties, e.g., \(\mathrm{cof} \bm{F} \tilde{v}_f\) that quantifies changes in area with respect to the fiber direction, \(\tilde{v}_f\), may be used instead of fiber compression to maintain stability under compression, e.g., in~\cite{ehretPolyconvexHyperelasticModel2007}.

Loss of ellipticity is often a precursor to formation of microstructure e.g., phase transition fronts, across which the state of strain and/or fiber orientation may be different~\cite{mollenkopf2025, liang}. In practice, in the presence of the matrix and fiber pre-stretch, loss of ellipticity of a single fiber did not appear to compromise the overall stability of the model. To show this, we compute the eigenvalues of the acoustic tensor following the standard procedure, outlined in~\cite{schroderVariationalApproachMaterially2005a} for example, for the uniaxial tension/compression results in the main paper. For all deformation states, we compute the eigenvalues in approximately 500 trial directions oriented randomly on a unit sphere and verify that they remain non-negative, indicating that the material remains stable despite the loss of ellipticity at the single fiber level. This result is shown in Figure~\ref{fig:min_eigen_vals}.

\begin{figure}
  \centering
  \includegraphics[width=0.8\textwidth]{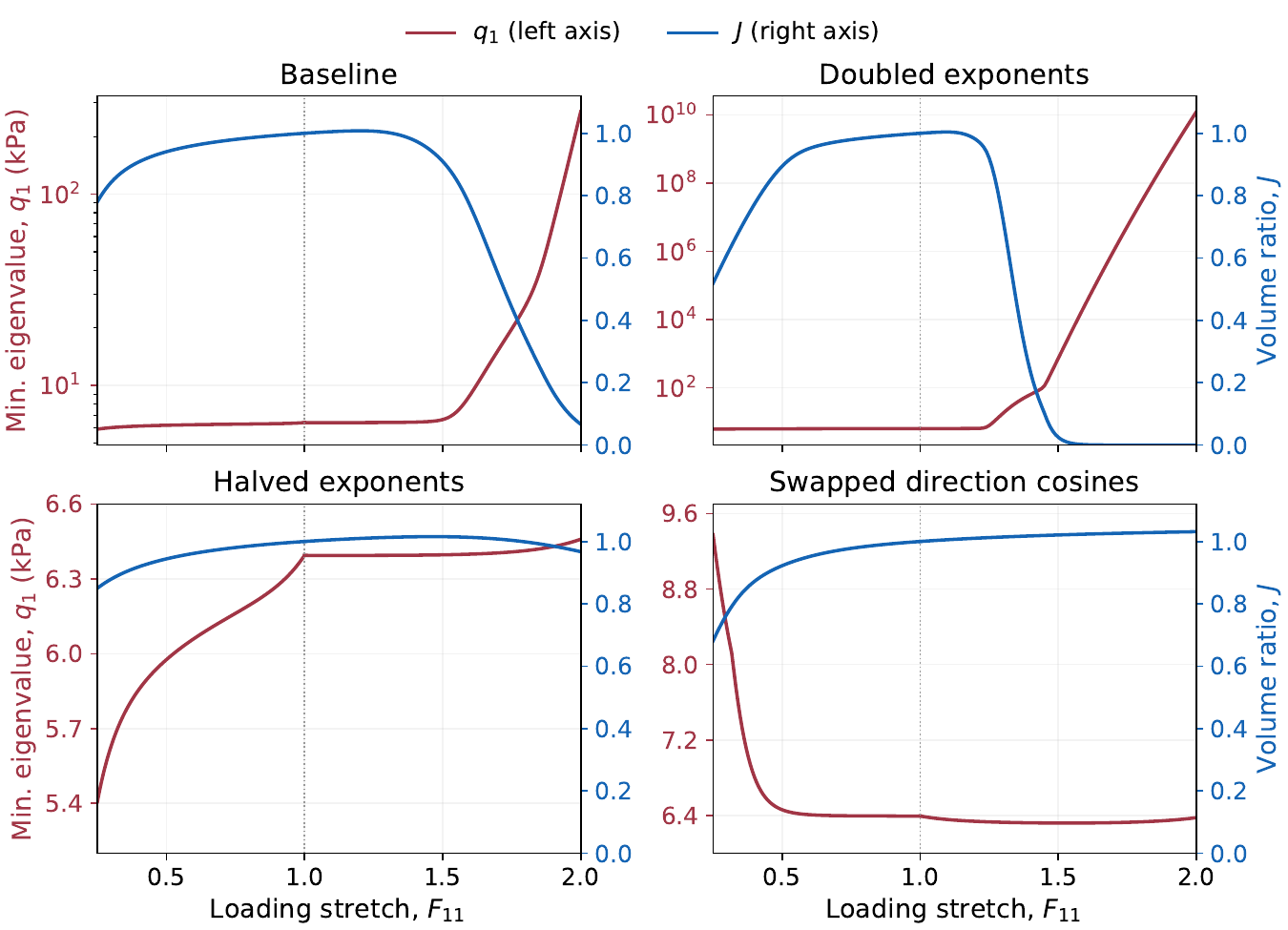}
  \caption{Smallest eigenvalues of the acoustic tensor for the uniaxial tension/compression test (Figure~\ref{fig:stress_volume} in main text). Non-negative eigenvalues indicate stability of the material despite the potential loss of ellipticity at the single fiber level.}
  \label{fig:min_eigen_vals}
\end{figure}

Finally, we note that loss of ellipticity due to extreme compaction of fibers is a consequence of modeling assumptions. In reality, fibers are more likely to buckle under such conditions rather than cause a loss of overall material stability, with additional stabilizing mechanisms such as inter-fiber contacts. This points to the need for more advanced material models to fully incorporate fiber compression in a realistic and stable manner.

\section{Isotropic Strain-Energy Function}
\label{sec:SI_isotropic_eng}
Here we present an isotropic equivalent of the anisotropic material model in the main paper. We use the same individual-fiber energy as in the main text, with a common stiffening exponent \(m=m_f=m_a=m_d\):
\begin{equation}
  \Psi_f = \frac{E_f A_f L_f}{2m-1}\left(\frac{1}{2m}\ifib{f}^{m} - \frac{1}{2} \log \ifib{f}\right)
\end{equation}

For isotropy, we assume (i) \(\alpha_i = 1/\sqrt{3}\), (ii) all fibers have identical material properties, i.e., \(E_f = E_a = E_d\), \(m_f = m_a = m_d\), and (iii) the eigenbasis of the deformation tensor forms the material coordinate system. The primary fiber directions and their corresponding invariants can now be written as follows in the eigenbasis:
\begin{equation}
  \begin{aligned}
    \tilde{a}_1 &= [1, 0, 0] \rightarrow \ifib{a_1} = \lambda_o^2 \lambda_1^2\\
    \tilde{a}_2 &= [0, 1, 0] \rightarrow \ifib{a_2} = \lambda_o^2 \lambda_2^2\\
    \tilde{a}_3 &= [0, 0, 1] \rightarrow \ifib{a_3} = \lambda_o^2 \lambda_3^2\\
    \tilde{d}_1 &= [1, 1, 1]/\sqrt{3}\\
    &\rightarrow \ifib{d_1} = \frac{\lambda_o^2}{3}(\lambda_1^2 + \lambda_2^2 + \lambda_3^2) = \frac{\lambda_o^2}{3}I_1 \\
    &= \ifib{d_2} = \ifib{d_3} = \ifib{d_4}
  \end{aligned}
\end{equation}

Here, \(\lambda_{1, 2, 3}\) are the principal stretches of the deformation tensor. With the axial fiber length equal to the diagonal fiber length, \(L_d\), divided by \(\sqrt{3}\), the contribution of the fiber system to the strain-energy density is given by:
\begin{equation}
  \begin{aligned}
    \Psi_{\mathrm{fs,iso}} &= \frac{2}{V_o}\left[\sum_{i=1}^3 \Psi_{a_i} + \sum_{j=1}^4 \Psi_{d_j}\right] = \frac{2}{V_o}\left[\sum_{i=1}^3 \Psi_{a_i} + 4 \Psi_d\right]\\
    & = \frac{2 E_f A_f L_d}{(2m - 1) V_o}\bigg[\sum_{i=1}^3 \frac{1}{\sqrt{3}}\left(\frac{1}{2m}\ifib{a_i}^{m}  - \frac{1}{2} \log \ifib{a_i}\right) \\
    &\qquad\qquad+ 4 \left(\frac{1}{2m} \ifib{d}^{m} - \frac{1}{2} \log \ifib{d}\right)\bigg]\\
    % simplify and get rid of constants in log
    & = \frac{\phi_f E_f}{(2m - 1)(4 + \sqrt{3})}\bigg[\sum_{i=1}^3 \frac{1}{\sqrt{3}}\left(\frac{1}{2m}\lambda_i^{2m}\lambda_o^{2m} - \frac{1}{2}\log \lambda_o^2\lambda_i^2\right)  \\
    &\qquad\qquad+ 2 \left(\frac{\lambda_o^{2m} I_1^m}{m \times 3^m} - \log \frac{\lambda_o^2 I_1}{3} \right)\bigg]\\
    & = \frac{\phi_f E_f}{(2m - 1)(4 + \sqrt{3})}\bigg[\frac{\lambda_o^{2m}}{2\sqrt{3}m}\sum_{i=1}^3 \lambda_i^{2m} - \frac{1}{\sqrt{3}}\log J \\
    &\qquad\qquad+ \frac{2\lambda_o^{2m} I_1^m}{m \times 3^m} - 2\log I_1 - (4 + \sqrt{3}) \log \lambda_o + 2 \log 3 \bigg]
  \end{aligned}
\end{equation}

Here, \(J = \det \bm{F}\) is the determinant of the deformation gradient and \(\phi_f\) is the fiber volume fraction from the main text. Overall, the resulting isotropic energy consists of a sum of Ogden-like principal stretch and Arruda--Boyce analogous invariant terms with logarithmic penalties.

\section{Parameter Calibration, Sensitivity Analysis, and Comparison}
\label{sec:SI_param_calibration}
\subsection{Parameter Calibration with FEMU}
The FEMU-based material parameter calibration used in this work is visualized in Figure~\ref{fig:calibration_schematic}. Here, we provide additional details and describe the implementation of each step.

\begin{enumerate}
  \item \textbf{Experiments} (from literature): Experimental data were collected from the five in-plane displacement-controlled biaxial protocols of Ross et al.~\cite{rossBiaxialMechanicalCharacterizations2019}: biaxial, pure shear in the circumferential and radial directions, and constrained uniaxial stretching in the circumferential and radial directions. We performed interpolation-based preprocessing on the digitized data to ensure data availability at the required deformation states.

  \item \textbf{Material Model} (FEA and AD): The octahedral strain-energy function, stress, and tangent are implemented in the FE solver. The strain-energy function is also implemented in an automatic-differentiation (AD) framework, JAX in our case~\cite{jax2018github}. For this calibration process, no pre-stretching was assumed. Additionally, the fiber volume fraction, \(\phi_f\), was set to 0.1.

  \item \textbf{FEA}: An initial admissible parameter vector, \(\bm{\Theta}\), is selected. The FE solver is used to obtain the deformation gradients for all test scenarios from Step 1 for this \(\bm{\Theta}\). The resulting deformation gradients are then used as input to the AD framework for volume ratio and stress calculation. A one-element brick model was used for all simulations to reduce the computational costs. The applied boundary conditions are set up so that the computed stretch pair, e.g., (\(F_{11}, F_{22}\)), follows the experimental trajectory. This is enforced through the use of an equivalent load-amplitude (ramp) function. We considered the out-of-plane direction to be traction free.

  \item \textbf{Parameter Update}: The deformation gradients are evaluated in the AD implementation to calculate the model stress, the stress-based loss, and its gradient with respect to \(\bm{\Theta}\). We did not consider the effect of the free surface in the gradient for this protocol. An optimizer updates \(\bm{\Theta}\), subject to the specified bounds and symmetry constraints. We used the Adam optimizer with weight decay and gradient clipping, but any other gradient-based optimizer could be used. The stress loss is calculated as \(\mathcal{L}_P = \frac{\langle(\bm{P}_{\mathrm{model}} - \bm{P}_{\mathrm{exp}})^2\rangle}{\langle\bm{P}_{\mathrm{exp}}^2\rangle}\), where \(\langle\Box\rangle\) denotes mean value, \(\bm{P}_{\mathrm{model}}\) and \(\bm{P}_{\mathrm{exp}}\) are the model-predicted and experimentally measured first Piola--Kirchhoff stresses, respectively. The losses are computed componentwise and later summed. The stress loss is supplemented with parameter-related penalty terms based on the specified thresholds and out-of-plane stress that should be zero for the current dataset.

  \item \textbf{Loop}: Steps 3 and 4 are repeated until the loss and parameter estimates satisfy the selected convergence criteria. We used an early stopping criterion to terminate the optimization if no significant improvement was observed over a predefined number of iterations. The FE solution is refreshed throughout the optimization with the current best estimates of the material parameters due to their interdependence. The FE solution may be refreshed at every iteration, though we found it sufficient to refresh every few hundred iterations to balance computational cost and convergence stability.
\end{enumerate}

\begin{figure}
  \centering
  \includegraphics[width=0.5\linewidth]{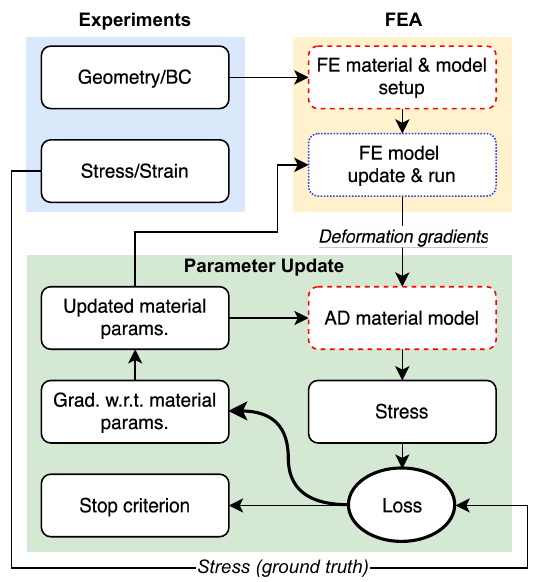}
  \caption{Schematic of the automatic-differentiation (AD)-assisted FEMU-based material parameter calibration procedure.}
  \label{fig:calibration_schematic}
\end{figure}
When the calibration protocol is applied to the current experimental dataset, we have used two procedural simplifications: (i) the dependence of the loss on the free surface, i.e., \(\lambda_3\), via parameter update has been ignored, and (ii) the deformation gradients were not updated at every iteration. The rationale for these simplifications is to reduce computational cost while maintaining reasonable accuracy in the parameter estimation. While it is likely that incorporating the total gradient in (i) would improve the convergence rate, it involves additional non-linear solution at every iteration. Our approach is also in line with prior appearance of this protocol, e.g., in~\cite{alhassaniehEfficientMaterialModel2025}. As for the second simplification, this is due to the empirical observation that, with other hyperparameters held constant, sparse update to deformation gradient is sufficient to achieve convergence. In fact, at a fixed compute, we find that this approach yielded better fit than the alternative of updating deformation gradient at every step.

The parameter estimates approached their final values within approximately 10 thousand iterations for the current dataset. The wall time for the entire calibration process was approximately 15 minutes on a MacBook Pro (M4 Pro CPU) laptop, most of which was spent on FE simulations and subsequent postprocessing to prepare the deformation gradients for the AD framework. The converged parameter values are tabulated in Table~\ref{tab:response_calibrated_parameters}. The \(R^2\) values for the calibrated model in various tests are given in Table~\ref{tab:r2-calibration} except for certain cases where the data were insufficient or not applicable due to constraints on the displacement. The fit metric shows some load scenarios are better represented by the calibrated model compared to others. The underlying reason is due to one or more of the following contributing factors: calibration procedure, experimental noise, limitations of the material model. In an exploratory study where we fit the material model to individual load case, the fit is far better than what's reported here for simultaneous fit. This suggests that the material model is capable of capturing the individual load responses well, but simultaneous fitting introduces compromises that prevent perfect agreement across all load cases. This is an expected tradeoff that arises from the need to balance accuracy across multiple loading scenarios simultaneously and present in other models, e.g., HGO, discussed in next section.

\begin{table}[!htbp]
  \centering
  \caption{Octahedral model parameters fitted to the five experimental protocols. The fiber volume fraction \(\phi_f=0.1\) and fiber prestretch \(\lambda_o=1\) were held fixed during calibration.}
  \label{tab:response_calibrated_parameters}
  \begin{tabular}{lrrrrrrrr}
    \hline
    Parameter & \(E_a\) & \(E_d\) & \(m_a\) & \(m_d\) & \(E_n\) & \(\nu\) & \(\alpha_1\) & \(\alpha_2\) \\
    Value & 18.79 & 23.91 & 7.66 & 14.97 & 20.89 & 0.47 & 0.92 & 0.38 \\
    Unit & kPa & kPa & \textemdash & \textemdash & kPa & \textemdash & \textemdash & \textemdash \\
    \hline
  \end{tabular}
\end{table}

\begin{table}[ht]
  \centering
  \caption{$R^2$ values for the FEMU direct calibration cases.}
  \label{tab:r2-calibration}
  \begin{tabular}{lcc}
    \hline
    Case & $R^2_{\mathrm{Circ}}$ & $R^2_{\mathrm{Rad}}$ \\
    \hline
    Biaxial
    & 0.959
    & 0.823 \\
    Pure Shear (circ. direction)
    & 0.475
    & 0.118 \\
    Pure Shear (radial direction)
    & 0.545
    & 0.945 \\
    Constrained Uniaxial (circ. direction)
    & 0.757
    & - \\
    Constrained Uniaxial (radial direction)
    & -
    & 0.957 \\
    \hline
  \end{tabular}
\end{table}

\subsection{Comparison to HGO}
The Holzapfel--Gasser--Ogden (HGO) model and its variants have been used to model aortic and mitral valve leaflets in studies of biaxial tissue mechanics, valve closure, and fluid--structure interaction~\cite{maoFullyCoupledValveSimulation2017,fengChordaeStructureMitralValve2018,aguileraAnnularDynamicsBarlow2021,aggarwalBayesianConstitutiveModelSelection2023,khaledianImagebasedSimulationMitral2025a}. Here we compare the proposed model to the HGO model in terms of their ability to capture the mechanical response of the tissue. To this end, we utilize the framework described earlier and five set of experimental data referenced in the main paper for calibration.

The HGO material model is described by five parameters, namely: the isotropic matrix stiffness \(c\), the fiber stiffness \(k_1\), the fiber nonlinearity parameter \(k_2\), the fiber dispersion parameter \(\kappa\), and the fiber orientation angle \(\gamma\). In the compressible variant we use (as implemented in FEBio), the compressibility is controlled by additional bulk parameter, \(K_o\)\cite{maasFEBioFiniteElements2012a}. These parameters are typically calibrated using experimental data to accurately capture the anisotropic mechanical response of the tissue. For heart valve tissue, a common strategy is to set \(\gamma = 0, \kappa = 0\) to reflect the aligned fiber structure and negligible dispersion. In our study, however, such assumption led to poor fit of the model. We therefore consider alternatives such as \(\gamma = 45^\circ\) (two fiber families about a symmetry line capturing the circumferential and radial direction) and nonzero \(\kappa\) to better capture the observed mechanical behavior. The results of the calibration are shown in Figure~\ref{fig:HGO_calib_variants}.

\begin{figure}[htbp]
  \centering
  \includegraphics[width=0.8\textwidth]{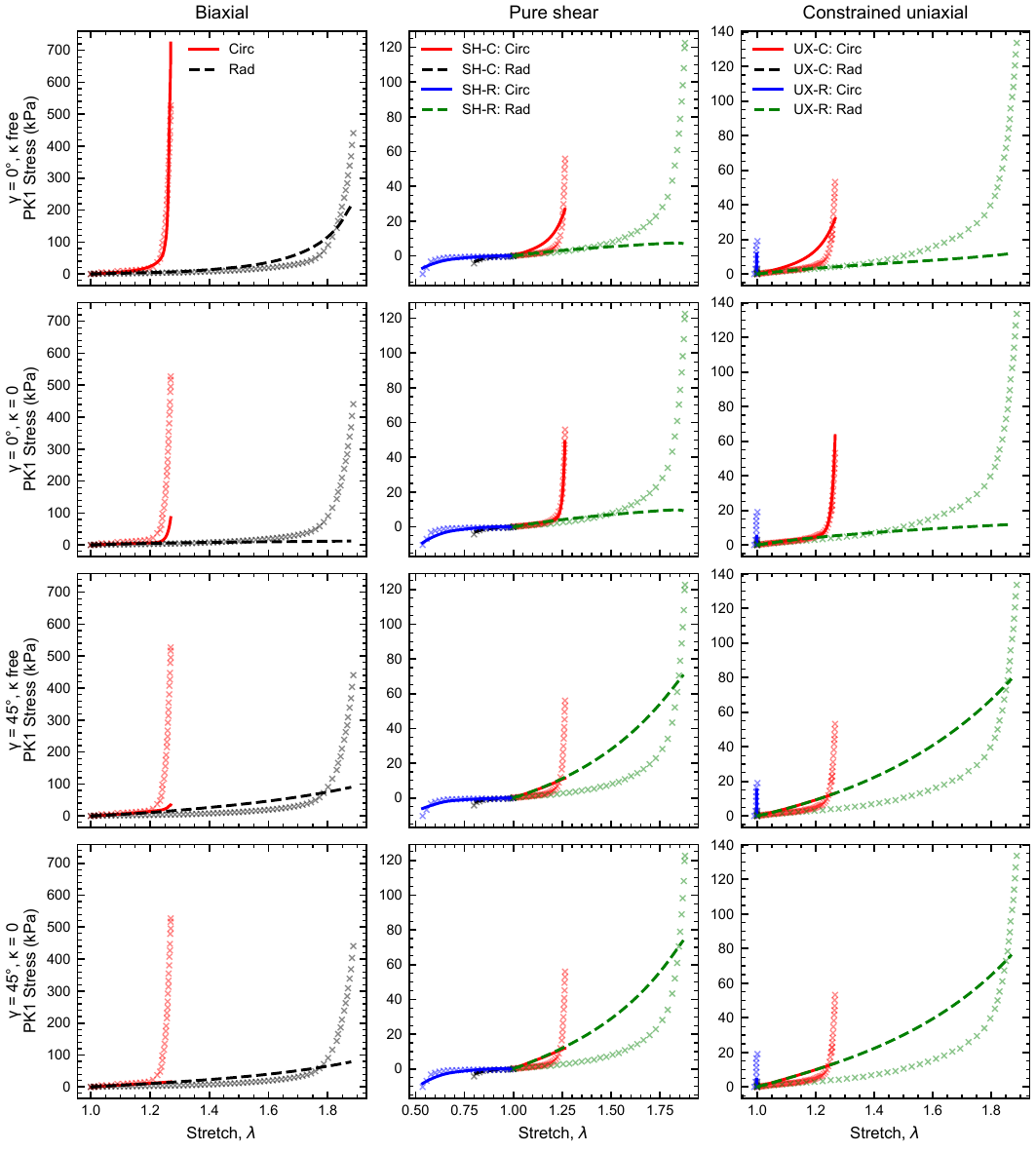}
  \caption{Calibration results for the HGO model with different parameter settings.}
  \label{fig:HGO_calib_variants}
\end{figure}

The HGO variant with aligned, nondispersed fibers (\(\gamma = 0^\circ\), \(\kappa = 0\)) fit the biaxial data poorly. Allowing fiber dispersion at \(\gamma = 0^\circ\) improved the biaxial fit and yielded the lowest joint calibration loss among the four HGO variants. The corresponding parameters used in the FEBio simulations were \(c = 5.30\,\mathrm{kPa}\), \(K_o = 135\,\mathrm{kPa}\), \(k_1 = 6.71\,\mathrm{kPa}\), \(k_2 = 8.26\), and \(\kappa = 0.142\). The proposed octahedral model provided better agreement with the experimental data across the five protocols considered jointly. In a simulation with these parameter values, HGO-based valve model showed significantly more billowing compared to the proposed octahedral model, indicating that the latter better constrains the leaflet deformation and more accurately represents the mechanical behavior observed in the experiments.

\subsection{Sensitivity Analysis}
The numerical optimization procedure used for parameter calibration involves minimizing a joint loss function that quantifies the discrepancy between the model predictions and the experimental data across multiple loading protocols. This procedure allows us to systematically explore the parameter space and identify the set of parameters that best captures the observed mechanical behavior of the tissue. The combination of available data, material model, and optimization framework, however leads to variations in converged parameter values and, subsequently, solutions to boundary value problems~\cite{ogdenFittingHyperelasticModels2004}. Here we quantify these for the octahedral model and HGO (\(\gamma = 0^\circ\), \(\kappa = \text{free}\) variant).

To assess the sensitivity of the model predictions to variations in the calibrated parameters, we perform multiple optimization runs starting from different initial guesses for the material parameters. We then examine the range of predicted stresses in the octahedral model across the ten lowest-loss runs to quantify the uncertainty associated with the parameter calibration. The resulting spread for the proposed octahedral model is shown in Figure~\ref{fig:sensitivity_analysis_octahedral}. The result indicates that while there is some variability in the predicted stresses, the overall mechanical response remains consistent across different parameter sets, demonstrating the robustness of the calibration procedure.

The corresponding spread for the HGO model is shown in Figure~\ref{fig:sensitivity_analysis_hgo}. Compared to the octahedral model, the HGO model exhibits negligible variability in the predicted stresses, even though the underlying fit metric remains poor. This indicates that the calibration framework with five sets of data is nearly constrained for HGO model with less number of free parameters, even though it does not translate to improved agreement with the experimental observations.

\begin{figure}
  \centering
  \includegraphics[width=0.8\textwidth]{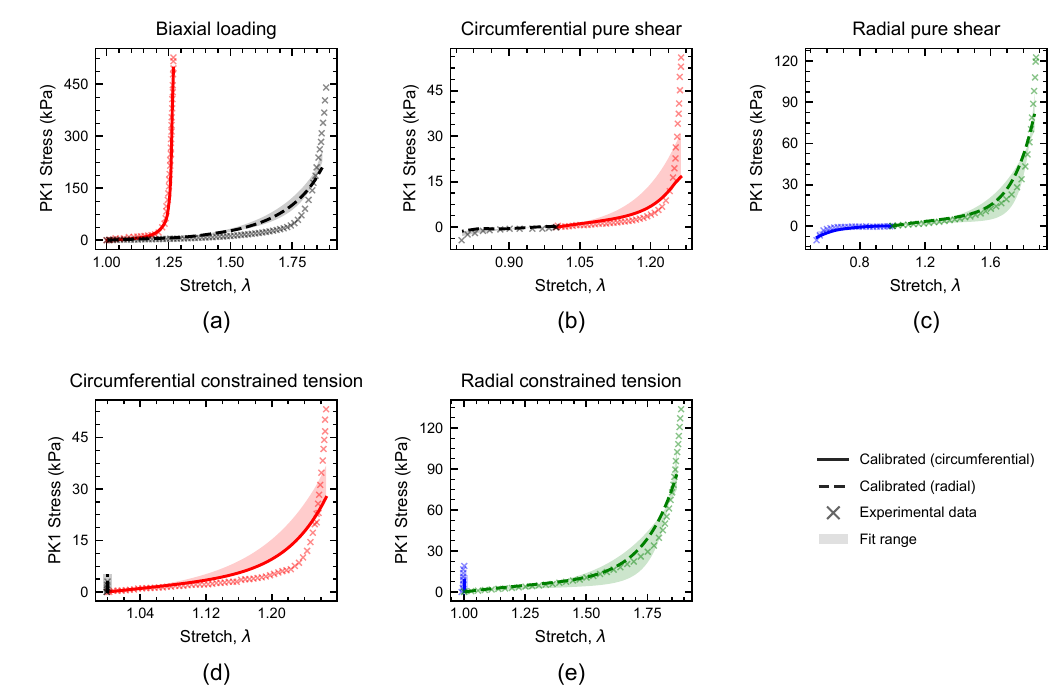}
  \caption{Stress predictions in the octahedral model from the manuscript calibration and the pointwise range across the ten lowest-loss runs from different initial material parameter vectors.}
  \label{fig:sensitivity_analysis_octahedral}
\end{figure}

\begin{figure}
  \centering
  \includegraphics[width=0.8\textwidth]{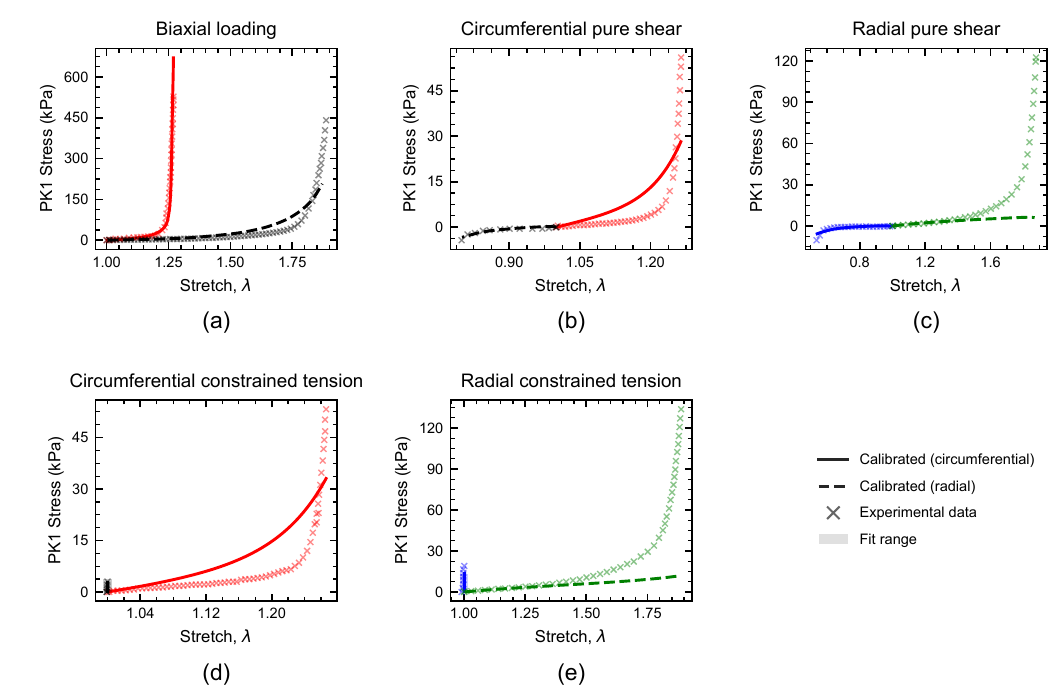}
  \caption{Stress predictions from the HGO model and the pointwise range (not visible due to negligible variability) across the ten lowest-loss runs from different initial material parameter vectors.}
  \label{fig:sensitivity_analysis_hgo}
\end{figure}
\par

\section{Reverse Poynting Effect}
\label{sec:SI_reverse_poynting_sshear}
Consider a simple shear deformation with shear strain \(\gamma\). For simplicity, we assume no pre-stretch and \(\phi_f = 1\). Assuming that the preferred fiber directions align with the global coordinate axes, the deformation gradient is given by:
\begin{equation}
  \bm{F} =
  \begin{bmatrix}
    1 & \gamma & 0\\
    0 & 1 & 0\\
    0 & 0 & 1
  \end{bmatrix} \rightarrow \bm{C} = \bm{F}^\text{T} \bm{F} =
  \begin{bmatrix}
    1 & \gamma & 0\\
    \gamma & 1 + \gamma^2 & 0\\
    0 & 0 & 1
  \end{bmatrix}
\end{equation}

The invariants for the fibers are given by:
\begin{equation}
  \begin{aligned}
    I_{a_1} &= 1\\
    I_{a_2} &= 1 + \gamma^2\\
    I_{a_3} &= 1\\
    I_{d_1} &= 1 + 2 \alpha_{1} \alpha_{2} \gamma + \alpha_{2}^{2} \gamma^{2} = I_{d_3} = K\\
    I_{d_2} &= 1 - 2 \alpha_{1} \alpha_{2} \gamma + \alpha_{2}^{2} \gamma^{2}  = I_{d_4} = L
  \end{aligned}
\end{equation}

This leads to the following stress contributions from the fibers
where \(\bm{S}_i\) denotes the stress contribution from the \(i\)-th fiber:
\begin{equation}
  \begin{aligned}
    \bm{S}_{a_1} &= \bm{0}\\
    \bm{S}_{a_3} &= \bm{0}\\
    \bm{S}_{a_2} & = \frac{E_a \alpha_2}{2m_a - 1}\left[(1 + \gamma^2)^{m_a - 1} - \frac{1}{(1 + \gamma^2)}\right] (\tilde{a}_2 \otimes \tilde{a}_2)\\
    \bm{S}_{d_1} &= \frac{E_d}{2m_d - 1}\left[K^{m_d - 1} - \frac{1}{K}\right] (\tilde{d}_1 \otimes \tilde{d}_1)\\
    \bm{S}_{d_2} &= \frac{E_d}{2m_d - 1}\left[L^{m_d - 1} - \frac{1}{L}\right] (\tilde{d}_2 \otimes \tilde{d}_2)\\
    \bm{S}_{d_3} &= \frac{E_d}{2m_d - 1}\left[K^{m_d - 1} - \frac{1}{K}\right] (\tilde{d}_3 \otimes \tilde{d}_3)\\
    \bm{S}_{d_4} &= \frac{E_d}{2m_d - 1}\left[L^{m_d - 1} - \frac{1}{L}\right] (\tilde{d}_4 \otimes \tilde{d}_4)\\
  \end{aligned}
\end{equation}

The overall stress is then calculated as:
\begin{equation}
  \begin{aligned}
    \bm{S} &= \Phi_f \left[\sum_{i=1}^3 \bm{S}_{a_i} + \sum_{j=1}^4 \bm{S}_{d_j}\right]\\
    &= \Phi_f \left[\bm{S}_{a_2} + ( \bm{S}_{d_1} + \bm{S}_{d_3}) + (\bm{S}_{d_2} + \bm{S}_{d_4})\right]\\
    &= \frac{\Phi_f E_a \alpha_2}{(2m_a - 1)}\left((1 + \gamma^2)^{m_a - 1} - \frac{1}{(1 + \gamma^2)}\right)
    \begin{bmatrix}
      0 & 0 & 0\\
      0 & 1 & 0\\
      0 & 0 & 0
    \end{bmatrix} \\
    &\qquad + \frac{2 \Phi_f  E_d}{(2m_d - 1)}\left(K^{m_d - 1} - \frac{1}{K}\right)
    \begin{bmatrix}
      \alpha_1^2 & \alpha_1 \alpha_2 & 0\\
      \alpha_1 \alpha_2 & \alpha_2^2 & 0\\
      0 & 0 & \alpha_3^2
    \end{bmatrix}\\
    &\qquad + \frac{2 \Phi_f E_d }{(2m_d - 1)} \left(L^{m_d - 1} - \frac{1}{L}\right)
    \begin{bmatrix}
      \alpha_1^2 & -\alpha_1 \alpha_2 & 0\\
      -\alpha_1 \alpha_2 & \alpha_2^2 & 0\\
      0 & 0 & \alpha_3^2
    \end{bmatrix}
  \end{aligned}
\end{equation}

The stress component of primary interest, \(S_{22}\), is then:
\begin{equation}
  \begin{aligned}
    S_{22} &= \frac{\Phi_f E_a \alpha_2}{(2m_a - 1)}\left(\frac{1}{(1 + \gamma^2)}\left((1 + \gamma^2)^{m_a} - 1\right)\right) \\
    &\qquad\qquad + \frac{2 \Phi_f E_d \alpha_2^2}{(2m_d - 1)}\left(\frac{1}{K}(K^{m_d} - 1)+ \frac{1}{L}(L^{m_d} - 1)\right)\\
  \end{aligned}
\end{equation}

To establish the reverse Poynting effect, it is sufficient to show that \(S_{22} = \sigma_{22} > 0\) for some \(\gamma\) and \(\sigma_{11} \neq \sigma_{22}\). The normal stresses are \(\sigma_{11} = S_{11} + 2 \gamma S_{12} + \gamma^2 S_{22}\) and \(\sigma_{22} = S_{22}\). Hence, typically \(\sigma_{11} \neq \sigma_{22}\) for a non-zero shear. Thus, the condition for reverse Poynting effect is \(S_{22} > 0\) for any non-zero \(\gamma\). In the expression for \(S_{22}\), the repeated pattern is \(\frac{1}{x}(x^p - 1)\) with \(x\) equal to \(1 + \gamma^2\), \(K\), or \(L\). Clearly, the expression is positive for \(x > 1\) when \(p > 0\).

Assuming \(m_a, m_d > 2\), the contribution from \(\bm{S}_{a_2}\) is always positive. In the remaining part of the expression, \(K^{m_d - 1} + L^{m_d - 1} - \frac{1}{K} - \frac{1}{L}\), the first two terms are convex for the relevant range of \(K, L > 0\). Therefore, \(K^{m_d - 1} + L^{m_d - 1} \geq 2\left(\frac{K + L}{2}\right)^{m_d - 1} = 2\left(1 + t\right)^{m_d - 1} > 0\), where \(t = \alpha_2^2 \gamma^2 \geq 0\). The other terms give \(1/K + 1/L = (K + L)/(KL) = 2(1 + t)/(KL)\). Overall, the lower bound of the expression is given by:
\begin{equation}
  K^{m_d - 1} + L^{m_d - 1} - \frac{1}{K} - \frac{1}{L} \geq 2\left(1 + t\right)\left[\left(1 + t\right)^{m_d - 2} - \frac{1}{KL}\right]
\end{equation}

The required condition is then \(\left(1 + t\right)^{m_d - 2} - \frac{1}{KL} > 0 \equiv KL \times (1 + t)^{m_d - 2} > 1 \). Substituting \(K \times L = (1 + t)^2 - 4 \alpha_1^2 t > 0\), we get:
\begin{equation}
  g(t) = (1 + t)^{m_d} - 4 \alpha_1^2 t (1 + t)^{m_d - 2}  - 1\overset{?}{>} 0
\end{equation}

For small shear such that \(t \to 0\), linearization about \(t = 0\) gives \(t (m_d - 4 \alpha_1^2) + O(t^2)\). This is positive for \(m_d > 4 \alpha_1^2\). Since \(\alpha_1^2 < 1\), the condition is satisfied for \(m_d > 4\). Similarly, at \(t \to \infty\), the expression approaches \(t^{m_d}\), and thus leads to reverse Poynting effect.

For the general case, \(\frac{\partial g}{\partial t} = (1+t)^{m_d-3}\left[m_d t^2 + \left(2m_d - 4\alpha_1^2(m_d-1)\right)t + m_d - 4\alpha_1^2\right]\). Assuming \(m_d > 4\), with \(\partial g/\partial t > 0\) at \(t = 0\), we require that the gradient does not cross zero for any permissible \(t\), i.e., the discriminant in \(m_d t^2 + \left(2m_d - 4\alpha_1^2(m_d-1)\right)t + m_d - 4\alpha_1^2 = 0 \) is negative. This leads to the condition:
\begin{equation}
  \alpha_1^2 < \frac{m_d(m_d - 2)}{(m_d - 1)^2} = 1 - \frac{1}{(m_d - 1)^2}
\end{equation}

If we set \(m_d = 4\), for instance, this leads to the condition \(\alpha_1^2 < 8/9\) for \(S_{22} > 0\) for all \(\gamma \neq 0\). Thus, the model exhibits reverse Poynting effect for a lower bound of \(m_a > 2, m_d > 4\) and \(\alpha_1^2 < 8/9\) under the current sign convention.

These sufficient conditions restrict the alignment of the diagonal fibers with the shear direction. However, these are sufficient but not necessary conditions for reverse Poynting effect. The necessary conditions are likely to be less restrictive.

\section{Additional Results}
\label{sec:SI_case_study_extras}
In this section, we present several additional results related to the material model calibration and mitral valve case study that complement the findings discussed in the main paper.

In Figure~\ref{fig:FA_uniax_volume}, we compare the volume changes predicted by the calibrated material model with experimental results for soft tissues (porcine myocardium) reported in~\cite{ehretInversePoroelasticityFundamental2017}. Compared to five-test calibration procedure in the main text, this involves uniaxial tension test data only. We consider two scenarios: stress-stretch from uniaxial tension test and joint calibration using stress-stretch and volumetric data. In both scenarios, we compare the predicted volume changes against the experimental measurements. The results show that both approaches capture the experimental stress-stretch behavior accurately (\(R^2\approx 0.99\)). However, when only the stress-stretch data is used for calibration, the fitted model under-predicts volume reduction observed during uniaxial tension. Due to the flexibility of the material model, however, when the experimental volumetric data is included in the calibration, the predicted volume changes align closely with the experimental observations, demonstrating the importance of incorporating volumetric information for accurate modeling of soft tissue behavior when possible.

\begin{figure}
  \centering
  \includegraphics[width=\linewidth]{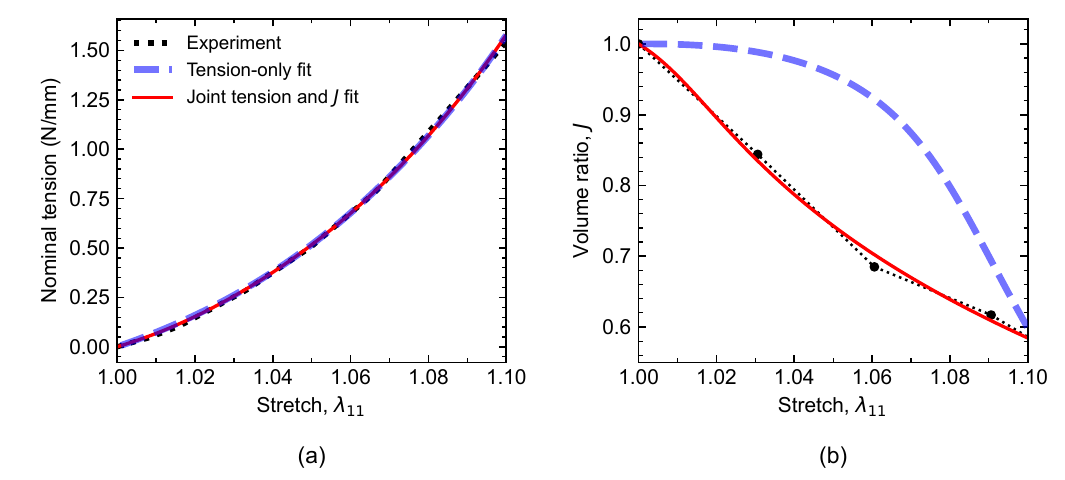}
  \caption{Uniaxial tension of porcine myocardium. (a) Nominal tension and (b) volume ratio \(J\) versus stretch. The black dotted curve shows the experimental data from~\cite{ehretInversePoroelasticityFundamental2017}, the blue dashed curve shows the tension-only fit, and the red solid curve shows the joint tension--volume fit.}
  \label{fig:FA_uniax_volume}
\end{figure}

Now we turn to the case study involving the mitral valve. Figure~\ref{fig:FA2_case_study_deltas} shows the relative differences in first principal strain and displacement magnitude due to localized perturbations in fiber orientation (see Figure~\ref{fig:local_reorient} in the main paper). The relative differences in the first principal values of the Lagrangian strain in Figure~\ref{fig:FA2_case_study_deltas}a mirror the stress-based result in the main paper as anticipated. The relative differences in the displacement are, on the other hand, smaller and more diffuse, likely due to the large rotation experienced by the valve leaflets during closure.

\begin{figure}
  \centering
  \includegraphics[width=\linewidth]{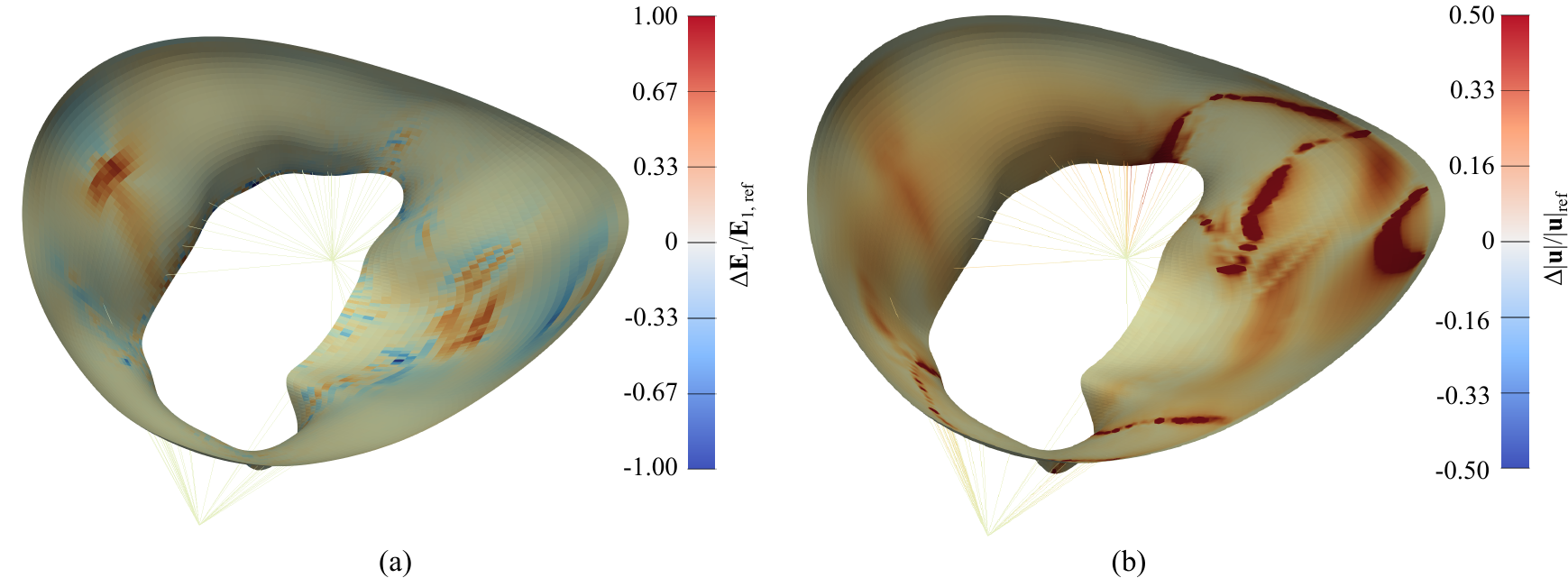}
  \caption{Relative differences in (a) first principal strain and (b) magnitude of the displacement field due to perturbation in fiber orientation in the anterior and posterior belly regions. The reference configuration corresponds to the unperturbed fiber orientation from Figure~\ref{fig:mitral_closure}a.}
  \label{fig:FA2_case_study_deltas}
\end{figure}

Figure~\ref{fig:FA3_extra_cases_deltas} compares the deformed configurations of the baseline case and a case with softer chordae. The constant in the Ogden model, \(c_1\), was reduced from 100 to 10 kPa in this case. This led to pronounced billowing in the valve leaflets compared to the baseline, as seen in panel (b) of the figure.

Finally, Figure~\ref{fig:FA4_extra_phase_diag_ciso} shows the phase diagram for the cubic anisotropic material, illustrating the mean principal strain and fraction of leaflet area in contact for different parameter combinations.

\begin{figure}
  \centering
  \includegraphics[width=\linewidth]{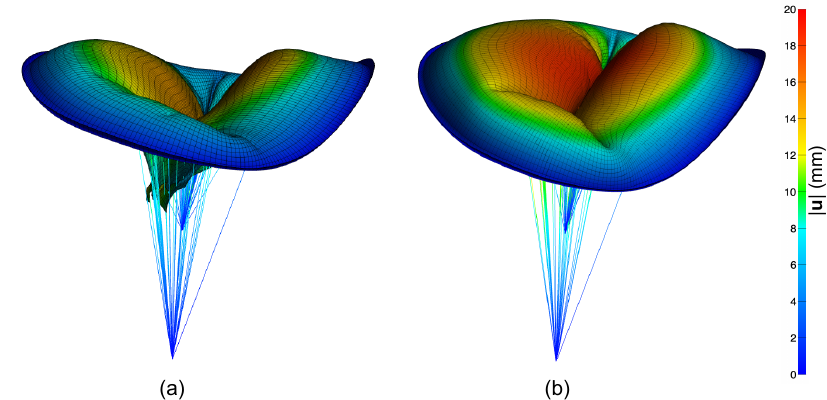}
  \caption{Deformed configurations for (a) baseline case and (b) a case with softer chordae. Color indicates the magnitude of the displacement field, capped at 20 mm.}
  \label{fig:FA3_extra_cases_deltas}
\end{figure}

\begin{figure}
  \centering
  \includegraphics[width=\linewidth]{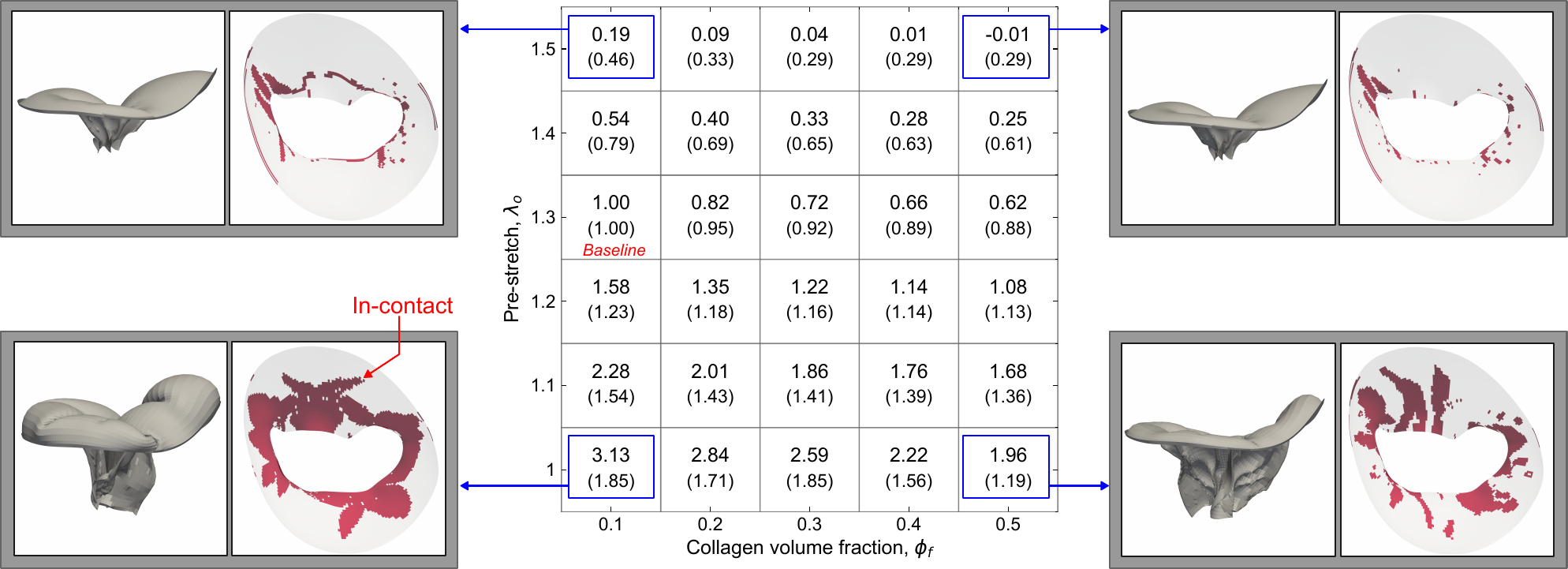}
  \caption{Phase diagram for cubic anisotropic material. The cell values represent the mean principal strain and fraction of leaflet area in contact (in parentheses). The results are all normalized by the corresponding results for \(\lambda_o = 1.3, \phi_f = 0.1\).}
  \label{fig:FA4_extra_phase_diag_ciso}
\end{figure}

% Bibliography embedded for submission; no external .bbl file is required.

\end{document}